# A Cyber Kill Chain Based Taxonomy of Banking Trojans for Evolutionary Computational Intelligence

**Dennis Kiwia [1], Ali Dehghantanha[1], Kim-Kwang Raymond Choo[2], Jim Slaughter[1]**

1: Department of Computer Science, School of Computing, Science and Engineering, University of Salford, United Kingdom

2: Department of Information Systems and Cyber Security, University of Texas at San Antonio, San Antonio, TX 78249, USA

D.Kiwia@edu.salford.ac.uk, a.dehghantanha@salford.ac.uk, raymond.choo@utsa.edu, slaughter.james@gmail.com,

## Abstract.

Malware such as banking Trojans are popular with financially-motivated cybercriminals. Detection of banking Trojans remains a challenging task, due to the constant evolution of techniques used to obfuscate and circumvent existing detection and security solutions. Having a malware taxonomy can facilitate the design of mitigation strategies such as those based on evolutionary computational intelligence. Specifically, in this paper, we propose a cyber kill chain based taxonomy of banking Trojans features. This threat intelligence based taxonomy providing a stage-by-stage operational understanding of a cyber-attack, can be highly beneficial to security practitioners and the design of evolutionary computational intelligence on Trojans detection and mitigation strategy. The proposed taxonomy is validated by using a real-world dataset of 127 banking Trojans collected from December 2014 to January 2016 by a major UK-based financial organisation.

## I. Introduction

In our current Internet-connected society, e-commerce and e-government (e.g. e-banking and e-payment systems) are becoming the norm in developing and developed nations. Such systems and services can and have been targeted by cyber criminals [1]–[6], and malware is a popular or common tool used by cyber criminals [7]–[9]. For example, a cyber-attack in South Korea reportedly saw 32,000 computers belonging to broadcasting organizations and banks infected with a malware that overwrote the Master Boot Record (MBR) [8]. Also, in 2013, the Crypto-Locker ransomware reportedly infected more than three million machines, causing more than 6 million USD worth of damages [10]. Malware can be broadly categorized in different aspects like network based and those that aren't network related [11], and further categorised into adware, spyware, virus, worms, backdoors, rootkits and Trojans [11]–[16].

Adware are designed with the purpose of displaying adverts in the computers they are running. Spyware as the term suggest, they spy on users without their knowledge and the information gathered can be used for all sort of purpose by the adversary [15]. Virus is a malicious code that spread through direct contact i.e. it must attach itself to another running program for it to reproduce/work [17]. Worms have the capability of self-replication and destruction i.e. does not depend on another program, often deletes data files from computers [11], [15], [17]. Backdoors also called trap doors, are malicious code embedded in applications or operating system with the intention of providing programmer access without requiring ordinary authentication method [11]. Rootkits are sets of software tools that enable an unauthorized user to gain administrative-level control of a computer system without being detected [11], [18], [19].

Trojans consists of two parts, server side which is usually small (few KBs) that runs on attacked host and client side that runs on attacker's console [15], [17]. Trojans have many ways of working, depending on the design they may facilitate/perform capabilities like backdoor, sniffing, spamming and so on [11], [15], [17]. There are many kinds of Trojans example Crypto-Locker that encrypts user files and request ransom to decrypt, Zeus a Trojan that steals victims online banking credentials to steal money from their account [17], these are also known as Banking Trojans.

Trojans are considered as one of the most persistent malwares that can evade conventional firewall and anti-virus capabilities over a significant period allowing adversaries to harvest sensitive information [20]. This paper will be focusing on banking Trojans, due to their capabilities to facilitate the hijacking or acquiring of online banking credentials and other sensitive information (e.g. credit card details), which are then sent through a backdoor to a Command and Control (C&C) infrastructure [21], [22].

The approach utilised by cybercriminals has made threat intelligence no longer a trivial aspect of defence for organisations due to the leveraging of Advance Persistent Threat (APT) [20], [23]–[25]. APT represents well-resourced and trained adversaries [20] that aims in acquiring high target and valued information. Conversional detection and prevention tools like firewall and anti-virus are often inadequate in detecting these attacks as they mostly concentrate on known vulnerabilities [2], [26]–[29]. Furthermore, incidence response methodology concentrate on after the fact effect of an intrusion [20].

While malware detection and mitigation research is not new, effectively detecting and mitigating malware remains challenging due to the constant evolution of malware and malware authors [30]. Cyber Kill Chain (CKC) is one of the most widely used operational threat intelligence models to explain intrusion campaigns activities [23], [31]. CKC is based on the kill chain tactic of the US military's F2T2EA (find, fix, track, target, engage and assess) [20], and contains seven stages/steps as shown on Figure 1 [20], [23], [31]–[34].

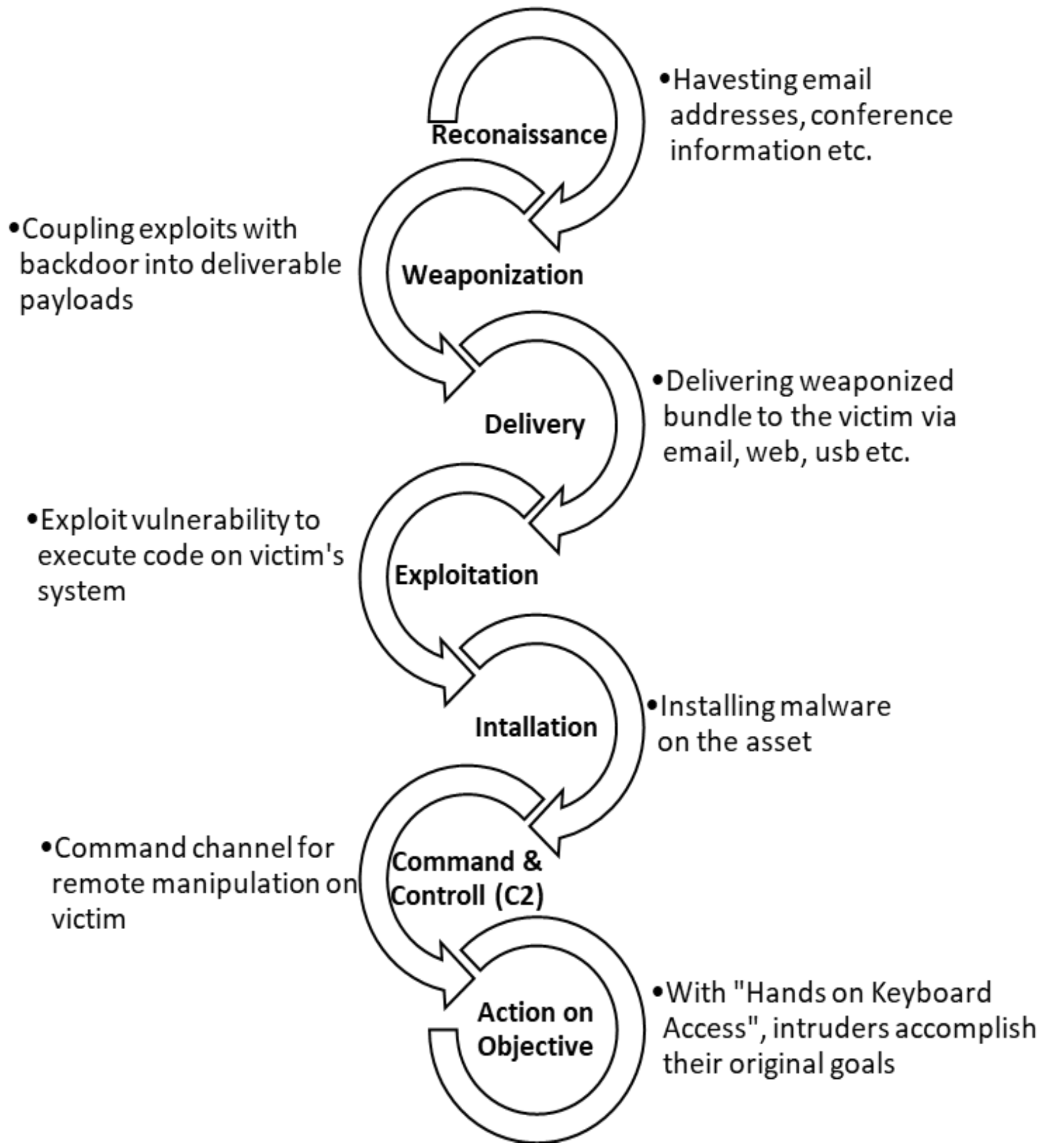

**Figure 1.** Lockheed Martin CKC steps

Reconnaissance includes the identification, selection and profiling of potential targets. In Weaponization, a cyber weapon is built by, say combining a Remote Access Trojan (RAT) with an exploit code (exploit kits), and efforts are made to minimize the risk of detection and investigation by the victim. In Delivery, the cyber weapon is transmitted to the victim(s) environment. Exploitation is the triggering / activating of the malicious payload in the target environment. During Installation, the cyber weapon allows the adversary to preserve its access and deliver more payloads to the victim environment. In Command and Control (C2), the adversary establishes communication with the compromised host(s). Finally, in actions on objectives, an adversary takes planned activities on target(s) (i.e. exfiltration of data) to achieve the intended goals.

Using threat model like CKC and aligning patterns of threat attacks to it, will allow researchers and security experts to speak in one language i.e. CKC. And in this language, we will be highlighting what to look for at each stage of the kill chain to support an intelligent-derived defensive and investigative tactics for analysts and experts in organisations to perform their day-to-day tasks against APT. For example: highlighting delivery features of Trojans, security teams will know what to expect and anticipate the countermeasure to prevent or minimise the likelihood of success at that stage. In providing data exfiltration tactics used by Trojans will assist in the analysis of traffics to identify malicious traffics identifiers and prevent data loss from organisations.

In this paper, we propose a CKC-based taxonomy for banking Trojan features which can be used to inform detection and mitigation strategies, such as those based on evolutionary computational intelligence. The practicality of the taxonomy is then validated against 127 banking Trojan samples collected from December 2014 to January 2016 from a real-world banking environment in the United Kingdom (see Appendix I). The Trojans were collected from multiple phishing campaigns that contained office attachments or a Multipurpose Internet Mail Extensions (MIME) encoded messages, which were reverse engineered to obtain the download URL for the executable file. Initial analysis on the executable was done using a tool called Static0.1 written by one of the authors (available at: https://github.com/slaughterjames/static/) and dynamic analysis by using https://malwr.com/ and https://www.hybrid-analysis.com/ to obtain C&C server address as quick as possible and other indicators of compromise (IOC) so as to block the malware to prevent damage to the company. To the best of our knowledge, this is the first Trojan taxonomy based on CKC.

This paper is organized as follows: Related work is given in Section II, in Section III the proposed Banking Trojans Features Taxonomy is covered, Section IV provides proposed Banking Trojans Defence Taxonomy and Section V covers Concluding Remark.

## II. Related Work

A taxonomy can be described as a classification that uses ontology with primary purpose of providing basis for processing, communicating and reasoning about the cyber-related threats [35]. Intelligence gathering has become one of the key aspect for successful defence against cyber-attacks, and taxonomies can play an essential part in supporting security experts to achieve this [35], [36]. In the following we review previous taxonomies produced by scholars on different variants of malware.

Weaver et al., proposed a taxonomy of computer worms based on the worm target discovery and selection strategies, carrier mechanisms, activation, and payloads [37]. Jacob et al., proposed a taxonomy for malware behaviours and divided malicious programs based on their behaviour into two main families namely simulation-based and formal detectors [38]. Lindorfer et al., suggested a taxonomy for malware evasive behaviours based on detecting environmental variables [39]. Ugarte-Pedrero et al., suggested a taxonomy for malware packers based on their runtime complexity measurement and code obfuscation method [40].

Karim et al., proposed a taxonomy for mobile botnets [41]. With emergence of mobile malwares, a taxonomy for mobile malware behavioural detection was proposed [42] followed by an android malware attack vectors taxonomy based on attackers modus-operandi [43]. Khattak et al., proposed taxonomies of botnet features and botnet detection and prevention techniques [36]. Dagon et al., developed a taxonomy of botnets structure depicting botnets key metrics and response strategies [44]. Gupta et al., proposed a taxonomy of various types of phishing attacks and defence solutions [45]. Rodríguez et al., proposed a taxonomy of POS (Point of Sale) RAM scraping malwares behaviour [46].

With each evolution of malware trends scholars have tried to produce a taxonomy for that trend to assist security experts to understand that trend so to create/implement appropriate defence. No taxonomy has been attempted to cover the notorious Banking Trojans in supporting security experts on the banking/financial industry sector. Further enhancing the taxonomy by aligning it with a well-known threat intelligence could be even more useful to security experts in different industries. This paper is aiming to cover that gap basing on effects banking Trojans have on banking/financial industry and its customers. This taxonomy will assist security experts to identify at which stage should the defence be thought from that will be both cost effective and useful for both the organisation and its customers.

## III. Proposed Banking Trojans Features Taxonomy

While reconnaissance is all about gathering information in preparation of an attack, in this paper we will purposely overlook this stage due to two main reasons. One there are many ways recon can be done which makes the subject space too huge and we do not want to concentrate on that. And second reason is essentially any attack can be a reconnaissance stage of another attack. However, Trojan attacks do utilise reconnaissance stage like any other cyber-attack. The features this paper will concentrate on are Weaponization, Delivery, Exploitation, Installation, C2 and Actions on Objectives to give readers the knowledge of Trojan's behaviours in these stages and how they accomplish their goals in each stage. Basing on validation by using the 127 Trojan samples collected from a real-world banking environment in the UK (see Appendix I), we present the proposed banking Trojans features taxonomy in Figure 2. Appendix II presents the mapping between the collected Trojans' features and our suggested Trojan features taxonomy.

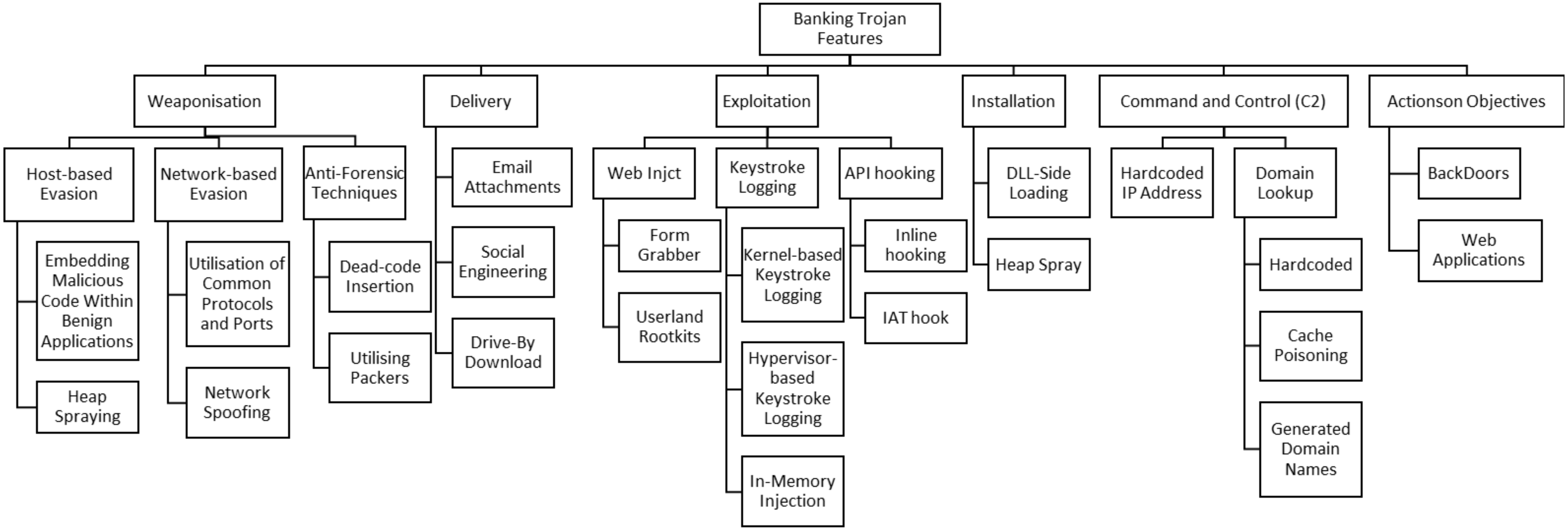

**Figure 2.** Taxonomy of Banking Trojans Features based on cyber kill chain.

# 1. *Weaponization*

Weaponization is an essential step to increase success of an attack by reducing opportunities for attack detection and limiting forensics investigators abilities to analyse detected threats[47]. Banking Trojans are employing variety of techniques to disguise detection on host-level and network-level and defeat cyber investigators.

## 1.1. Host-based Evasion

Host-based cyber defence is mainly relied on Anti-Virus (AV) and end-point security solutions installed on the host machines, which scans for known virus signatures or detect and limit malicious behaviours [48], [49]. Banking Trojans are using different techniques to install themselves on a host while remain undetected as follow.

***Embedding Malicious Code Within Benign Applications:*** Embedding a malicious code within a benign payload (application) using techniques such as process hollowing [50]–[53] would allow a malicious program to run within a known good process space and evade AVs detection even if the malware signature exists in the AV database [54], [55]. Adversaries that embed their executables within benign applications will infect the host next time that the benign app is called [49], [55], [56]. Client application data files such as Adobe Portable Document Format (PDF) or Microsoft Office documents are often used in this stage [20]. Out of 127 banking Trojans analysed in this research (shown in Appendix I), 96% utilised Microsoft Office and 4% utilised PDF.

***Heap Spraying;*** in this technique, malicious (shell) codes are divided into several pieces that are loaded in different locations on the heap memory so there is no single chunk of data mapped to any AV's signatures [56]. This technique not only reduces the chance of attack detection but increases the possibility of successful attack as adversary does not need to know the exact shell-code address in the heap but piece together shellcode elements using pre-defined gadgets[57]. As an example, Trojan 29 (see Appendix I - MD5 hash of ab40142988527fe6ce585a9fdfce56ca) has leveraged this technique to avoid detection. A potential jump to a section in a memory can be observed by memory analysis, a malware will have an instruction to execute a shell code in a specific offset e.g. "0x13ea60 (67): [*] Executing shellcode at offset: 0x1dbd8".

## 1.2. Network-based Evasion

Firewalls and Intrusion Detection/Prevention Systems (IDS/IPS) are the most common forms of network protection mechanisms. Similar to any other protection mechanism, the network perimeter defence solutions are having their own limitations i.e. a malicious document attached to an email is likely to bypass IDS while an executable would be subject to heavy scrutiny [54], [58]. Banking Trojans are using different techniques to avoid network-based detection as follow:

***Utilisation of Common Protocols and Ports;*** in this technique, the Trojan only utilises very common protocols and ports to conduct its malicious intents. Most Trojans are using protocols like HTTPS, DNS, and HTTP on ports such as 443, 53 and 80. This technique reduces the chance of port being restricted on the network perimeter and the communication being detected as malicious [36]. With the use of common ports, Trojans can communicate with their C2s, download payloads, and upload exfiltrated data. Analysis of network traffics using sniffer like wireshark can assist in identifying malicious ports and reviewing payloads associated with the traffic for potential data exfiltration.

***Network Spoofing;*** Spoofing is the art of deceiving someone by impersonating origin of a malicious content as if it is coming from a known good and reliable source. Symantec reported 74% of spam campaigns used real companies domain in the sender's address [59], [60]. Most banking Trojans obfuscate their origin to well-known companies' domains which are local to their victims (i.e. using British Telecom (BT), and Royal Mail for targets within UK) to avoid being blacklisted on network perimeters, detected and/or identified by any other means including by the end users themselves [61]–[68].

### 1.3. Anti-Forensics Techniques

Attackers not only evade detection mechanisms but also they will try to make it difficult for forensics investigators to understand their intentions [69], [70]. Malwares can exploit systems in ways to make it difficult to be realised or investigated easily. For example: Android malwares can perform a Cross App Exploitation whereby a genuine app is exploited by another app [71] and in scale-free networks (SFNs), nodes running similar software (monoculture) can spread malware among them due to similar vulnerabilities in software [72] both approach makes it difficult for investigators to identify the source of attack.

Apart from techniques like cross app exploitation and exploitation of the monoculture, at the malware level tools that are used by legitimate programmers to protect their programs from tampering or deter reverse engineers can also be used to pack and obfuscate malwares to throw investigators off track [29], [55], [73]. One of the most common anti-forensics technique used at the malware level is obfuscation. Obfuscation is a deliberate act of disguising codes to make investigation more difficult. Similar to other malwares Banking Trojans are employing different obfuscation techniques as follows [29], [74]:

***Dead-Code Insertion;*** this technique involves adding ineffective instructions to a programme to change its appearance but maintain its behaviour [29], [73], [74]. It could be as simple as adding "NOP" instructions in the malware code or putting hundreds of code lines which are never executed or serve no purpose [57], [73]. This technique is quite often used and 76 of Trojans analysed have indicated utilisation of this technique (see Appendix II). While this technique could be circumvented by removing ineffective instructions prior to analysis, detecting those instructions is quite time consuming[26], [74]–[78].

***Utilising Packers;*** in this technique the malware encrypts its main body and only include a decrypting module which decode encrypted instructions at run time  [29], [73]. This technique creates difficulties in static code analysis and code reversing [37] as investigator should either extract instruction from memory dumps or write/find a decryption tool [29], [55], [79]. Moreover, parkers change malwares signature hence, makes the Trojan invisible to signature-based detection technologies [31], [40], [42], [48].

## *2. Delivery*

Once adversaries have completed Weaponization stage they should find a way to deliver their malicious payload to intended targets [36], [80]–[82]. The most common methods of malicious payload delivery by banking Trojans are email attachments, social engineering and drive by download [17], [20], [23], [31].

### 2.1. Email attachments

Email communication is heavily utilised in most organisations and often associated with an office document or PDF attachment. These attachments may contain malicious codes in the form of macros (Microsoft Office documents) or Java Scripts (PDF files) [58], [83].  Once a user enables macros in an office document it may download a payload that contain a Trojan. Similarly if a PDF is viewed by a privileged user, a JavaScript can be automatically launched to run a malicious shellcode [27], [84]. Since anti-virus software often fail to detect these hidden malware and there is a good chance for users to run the file this mechanism becomes highly effective for delivering Trojans [54], [58], [83].

### 2.2. Social Engineering

Attackers that uses banking Trojans like Dridex, Zeus and Spyeye widely explored social engineering to prey on user's weaknesses, hoping the recipient will press a link to open a malicious web page or an attachment with embedded malicious code [54], [58], [83]. According to Symantec report [85] on Dridex campaign attackers have sent 271,019 disguised as financial emails e.g. invoice, orders and receipts between Nov 2015 to Jan 2016  to deliver their Trojan. This is a form of social engineering to lure a victim to open the attachment thinking it's something of importance but the attachment contains a malware that is activated through macros or JavaScript embedded on the document[86]–[88]. In our dataset, we found 122 Trojans that are delivered using office documents and 5 delivered using PDF files (see Appendix II).

### 2.3. Drive-by Download

This mechanism utilises active contents such as a JavaScript or ActiveX [36] to make the users knowingly (by authorizing to run the active content) or unknowingly download attackers malicious contents while browsing a compromised or specially crafted web page [17], [36]. Adversaries may include the malicious code in a seamless object like an advert or a widget and eventually lure a victim to download a Trojan without intending to do so [17].

## *3. Exploitation*

After delivery of the malware, intruders' code should be triggered on the target machine by running the malicious application or exploiting a system vulnerability [20], [23], [31]. A successful exploitation may lead to exfiltration of private information, injection of code into web applications, log keystrokes, steal passwords, steal cookies or download other modules that may perform intended malicious activities. Banking Trojans are utilising a variety of techniques for exploitation which can be categorised into Web injection, key stroke logging and API hooking.

### 3.1. Webinject

Trojans are equipped with a functionality called web-inject [22], that can silently modify a webpage on the infected victim's machine to intercept private credentials such as username, password and even 2nd factor authentication credentials [21], [22]. Two main features are mostly utilised for web injection by banking Trojans as follow:

***Form Grabber*:** this is one of the famous techniques deployed for example by SpyEye Trojan, to manipulate and inject arbitrary contents into data transmitted between an HTTP(s) server and a client browser [21], [22]. The module is placed between the browser rendering engine and the HTTP(S) API function, so that the Trojan has access to decrypted data even if an encryption is utilised (e.g. SSL). In this mechanism, the Trojan presents the victim browser with a web page which is almost identical to the requested online banking site, while web form fields contents such as username and password are intercepted by the Trojan.

***Userland Rootkits*;** In this technique the Trojan uses communication API hooking to inject its malicious code during initialization of victim's web browser and intercepts and manipulates web traffic [21]. Zeus v.2 is a well-known Trojan leveraging this technique by manipulating WININET.DLL library in loading of Internet Explorer by hooking high-level API communication functions such as HttpQueryInfoA, HttpSendRequestA, InternetReadFile, InternetReadFileExA, etc., in user-mode. Therefore, the Trojan can conveniently intercept data before it gets encrypted in sessions secured by HTTPS[21].

### 3.2. Keystroke Logging

In Keystroke logging an adversary covertly records user's keystroke as they are being typed either through a software program or a hardware device or even by monitoring electromagnetic emissions. Deployed through variety of methods, malware programmers often leverage software key loggers as a payload or a client-side exploit [89], [90]. Software based keystroke logging can be implemented in kernel, hypervisor or in memory.

***Kernel-based Keystroke Logging;*** this technique requires privileged access to the victim machine since the malware should run as root or system administrator. This type of keystroke logging is facilitated by kernel-mode rootkits, where by the rootkits modify the kernel code (for example system calls) or kernel data to change the kernel behaviour in order to enforce stealthy capabilities to hide malicious activities i.e. kernel level keystroke logging [91]. Keystroke logging done at this level are immune to techniques that reveal user-mode activities [91], [18], [19].

***Hypervisor-based Keystroke Logging;*** This type of Trojans resides on hypervisor level which is "Ring -1" [91] lower than kernel (i.e. "Ring 0"). Therefore, these Trojans are much stealthier giving more control to attackers. Virtual Machine Based Rootkits (VMBR) install a virtual machine monitor (VMM) underneath the existing operating system that facilitates hypervisor keylogging by Trojans.

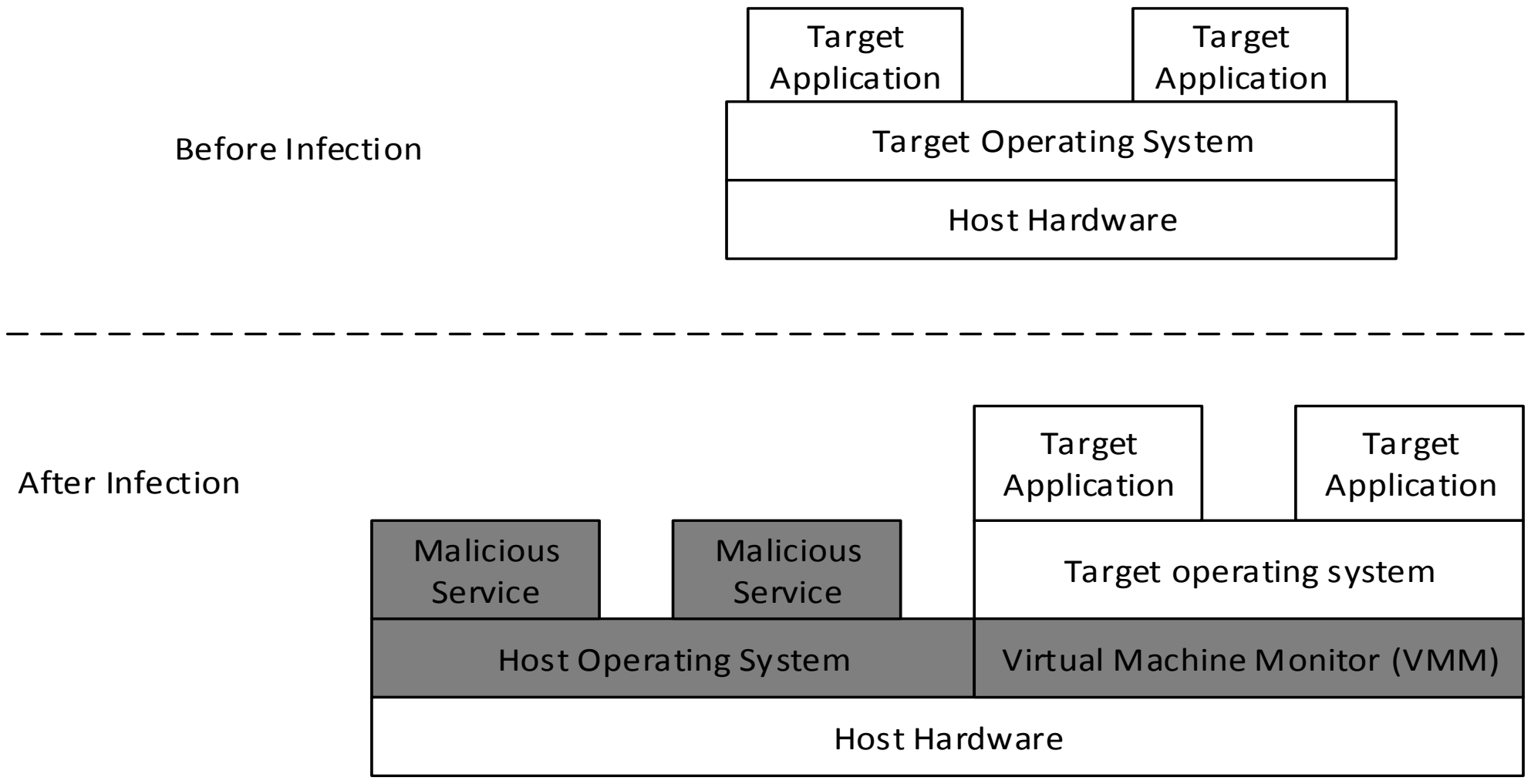

**Figure 3.** System structure of machine before and after infection of VMBR.

Once a machine is infected the malware can modify the VMM's emulated network card and log all network packets or use virtual-machine introspection to trap instructions and reconstruct data from the target system whilst being invisible to the target OS[18]. Since virtual-machine introspection can trap all SSL socket write calls, clear text data can be logged before being encrypted [18].

***In-Memory Injection;*** this keylogging mechanism can be achieved by patching memory tables or injecting directly into a process memory. Direct Memory Access (DMA) technique is usually used for injecting a key logger into a process [92], [93]. As memory injection happens at user level, attacker only requires similar privilege as injecting process for successful deployment of a keystroke into the process memory [90].

## 3.3. API Hooking

API hooking is a technique by which the behaviour and flow of applications can be influenced and modified through inserting memory break point and JMP (jump) instructions. API functions manipulations on system libraries such as Kernel32.dll, advapi32.dll and ntdll.dll can provide a privileged access to attackers [21]. Inline hooks, and Import Address Table (IAT) hooks are the main API hooking techniques implemented by different banking Trojans.

***Inline Hooking:*** this is a common Windows API hooking technique that replaces the byte code of an API function with a code redirection instruction to a code section controlled by the Trojan [21]. The 5 bytes at the beginning of every Windows system function for hot patching provides needed space for Trojan to implement this technique [21]. The 5 bytes will be replaced with NOP sled (No-operations slide) that will perform unconditional jump to a code section that contains Trojan code. Analysis of a memory dump from a compromised machine using volatility framework on Kali with command like "vol.py apihooks" can reveal in line hooking done by Trojans.

***Import Address Table (IAT) Hooks:*** whenever a Windows loader loads a PE (Portable Executable) file an IAT is filled with the virtual address of all functions or variables that are called or imported by the executable (DLL) [21], [94]. An IAT hook, overwrites the original destination of an imported API function and points it to the attacker code. Therefore, executing the application would load and run Trojan codes as well i.e. will activate the Trojan on the destination/host device.

## 4. *Installation*

In this stage the adversaries try to extend their access to more systems and compromise more nodes. Trojan developers often use tactics that are not easily detectable like DLL side loading and Heap Spray for a successful installation [20], [31].

### 4.1. DLL Side-Loading

Windows allows applications to load Dynamic-Link Libraries (DLLs) by either specifying full path of the DLLs or using DLL redirection, or utilising a manifest [95]. If none of these locate the DLL, then Windows will perform a search through predefined directories like WinSxS [95]–[97]. Attackers abuse this feature by putting their malicious DLLs in higher priority locations than original location of benign DLLs, hence the application would load the malicious DLLs instead [95]. With the use of Windows' Side-by-Side (WinSxS or SxS) a Trojan can load any malicious DLL from i.e. %WINDIR%\WinSxS (e.g. C:\Windows\WinSxS\x86_microsoft.windows.common-controls_6595b64144ccf1df_6.0.9600.16384_none_a9f4965301334e09) [95]. Due to the fact, most of this libraries are white listed in the National Software Reference Library (NSRL) database, most end-points will use this database to reduce false positives in their detection and since the execution of these libraries is done in the memory, this allows malicious attackers to perform more persistent attack with this technique as the detection will be ignored by most end-point protection software [95].

### 4.2. Heap Spray

This technique increases the chance of successful attack because attackers do not need to know the exact location of their malicious code in heap [57]. The heap spraying attack insert as many malicious code blocks as possible into the heap [57]. This technique usually is carried out in two phases i.e. NOP-sled and shell code, the NOP-sled reduces the chance of detection [28], [56], [57], [98]. Heap spraying usually uses a large amount of memory as it uses string object of JavaScript, the technique are mostly commonly used to compromise web browser and also on malicious PDF [56], [57], [98]. On Trojan "Trojan 29" this was observed clearly, when the binary reached address (0x13ea60) of its execution there is an instruction to execute shell code in the offset 0x1dbd8 "0x13ea60 (67): [*] Executing shellcode at offset: 0x1dbd8".

## 5. *Command and Control (C2)*

This is a mechanism by which a malware registers to its Command and Control (C&C) server [20], [23], [31], [36], [81]. Malware are registering on a C&C domain to receive commands or upload exfiltrated data. APT adversaries will make use of techniques of confidentiality and privacy such as utilisation of relay networks to ensure their anonymity to avoid prosecution [99]. A Trojan may find its C&C server through hard coded IP addresses or through domain lookup.

### 5.1. Hard Coded IP Address

This mechanism provides the C&C address by means of static IP address coded in the malware binary or by means of seeding [36]. In seeding the programmer provides the Trojan with an initial list of active peers in the botnet which are hidden anywhere on the infected machine with an elusive name to make it difficult to be detected. This mechanism is considered primitive since reverse engineering of the Trojan can easily lead to retrieval of the IP addresses or may even lead to detecting the actual botmaster.

### 5.2. Domain Lookup

The use of domain name system (DNS) is also very common among banking Trojans [100]. Unlike IP addresses, domain lookup provides stealthier tactical advantage to the malware developers due to its more complex nature. This mechanism has several advantages from the attacker's perspective such as:

- Any IP address can be linked to the domain name by the attacker, hence defeating IP blocking defence solutions.
- Completely blocking all the IPs linked to a domain name can be almost impossible as the attacker may reroute the traffic through several bots (stepping stones) before reaching the true Botmaster.
- Taking down a domain is complicated as it requires a lot of formalities.

Attackers have several options to include a domain name in their Trojans as follow:

***Hardcoded;*** Pre-defined domain names are hard-coded in the strings of the binary like in the case of IP address [36]. Attackers may have obtained different domain names that the victim is often visiting and try to host their Trojans on those or similar domains. Memory analysis can provide details of domains hardcoded on the malware, as this becomes visible when the malware is executed. With such information, you can blacklist the domain to protect your infrastructure.

***Cache Poisoning;*** attackers can magnify their malicious response using either web cache or even browser cache of a single user. If the response of websites is cached, attackers can manipulate the cached responses and redirect the victim to a different location (DNS cache poisoning attack) [31], [101]. This mechanism may produce different domain names each time the malware restarts and creates more polymorphic behaviour hence a stealthier malware.

***Generated Domain Names;*** The domain name can be dynamically generated by an algorithm (Domain Generation Algorithm) that is known to the attacker and the malware[36], [82]. Unlike cache poisoning this mechanism produces random domain name depending on the algorithm which also creates more polymorphic behaviour of the malware and makes it even stealthier [80], [82]. The difference between this technique and the cache poisoning attack is that in the cache poisoning the malware makes use of the domain cached in victim DNS while in this mechanism the malware produces a random domain name [31].

Out of 127 analysed Trojans 22 had IP addresses hardcoded while 13 utilised DNS only while the rest were using a combination of both IP and DNS for C2 (see Appendix II).

## *6. Actions on Objectives*

After an attack has completed all other steps in the kill chain what is left is to accomplish its objectives which could be data exfiltration and/or system disruption [20], [23], [31]. Banking Trojans have mostly concentrated their objectives in exfiltrating sensitive data from their victims for financial gain. Different banking Trojans utilises different technics to accomplish this but in high level this is done through a backdoor, FTP (File Transfer Protocol) or even Web and mobile apps. Banking Trojans like Carberp had gone to a level of implementing randomly generated cryptographic cipher to encrypt data obtained from the victim while sending back to its C&C [31], [102], [103], this way it ensures the payload is not blocked by any control such as data loss prevention devices.

### 6.1. Backdoors

Attackers often install a secret exit that allows them to send, receive or control victim machines remotely which is known as backdoor [45], [48]. Backdoor can be described as a feature or defect of a computer system that allows surreptitious unauthorized access to data [3], [104].

HTTP-based backdoors are very common in banking Trojans, due to their easiness in bypassing connection restrictions and detection. Moreover, the POST and GET commands are of the utmost interest when it comes to data exfiltration, both by receiving and sending data [33], [46]. 112 out of 127 Trojan samples in our dataset make use of backdoors for C&C communication via GET/POST HTTP requests (see Appendix II). Other than file upload, backdoors provide other means or advantages to attackers like remote login or even in-depth reconnaissance without users' knowledge[10], [31], [43]. Analysing web traffic with sniffer like wireshark you can observe payloads associated with GET/POST command and identify if data exfiltration is taking place and what kind of data is being exfiltrated.

## 6.2. Web Application

Trojans use MITB (Man-In-the-Browser) or MITM (Man-In-The-Middle) techniques to extract data [21], [105]–[107]. In MITB a Trojan redirects the victim to a phishing website that is controlled by the adversaries to harvest credentials such as username, password etc. [105], [106]. In MITM the adversary goes further by intercepting communications from victims and responses from server and establishing an interactive process for collecting users data [107], [108].

## IV. Proposed Banking Trojans Defence Taxonomy

In developing a banking Trojans defence taxonomy, we consider three dimensions, namely: detection, prevention and remedial actions [109]. The first level of our taxonomy comprises different types of defences, and the second level is constructed based on how a specific defence type can be applied. Detection techniques can run at the host or at the network [110], while prevention techniques are classified as host-based, network-based and user training. Finally, remedial actions are divided into defensive and offensive activities; each linked to different techniques [36].

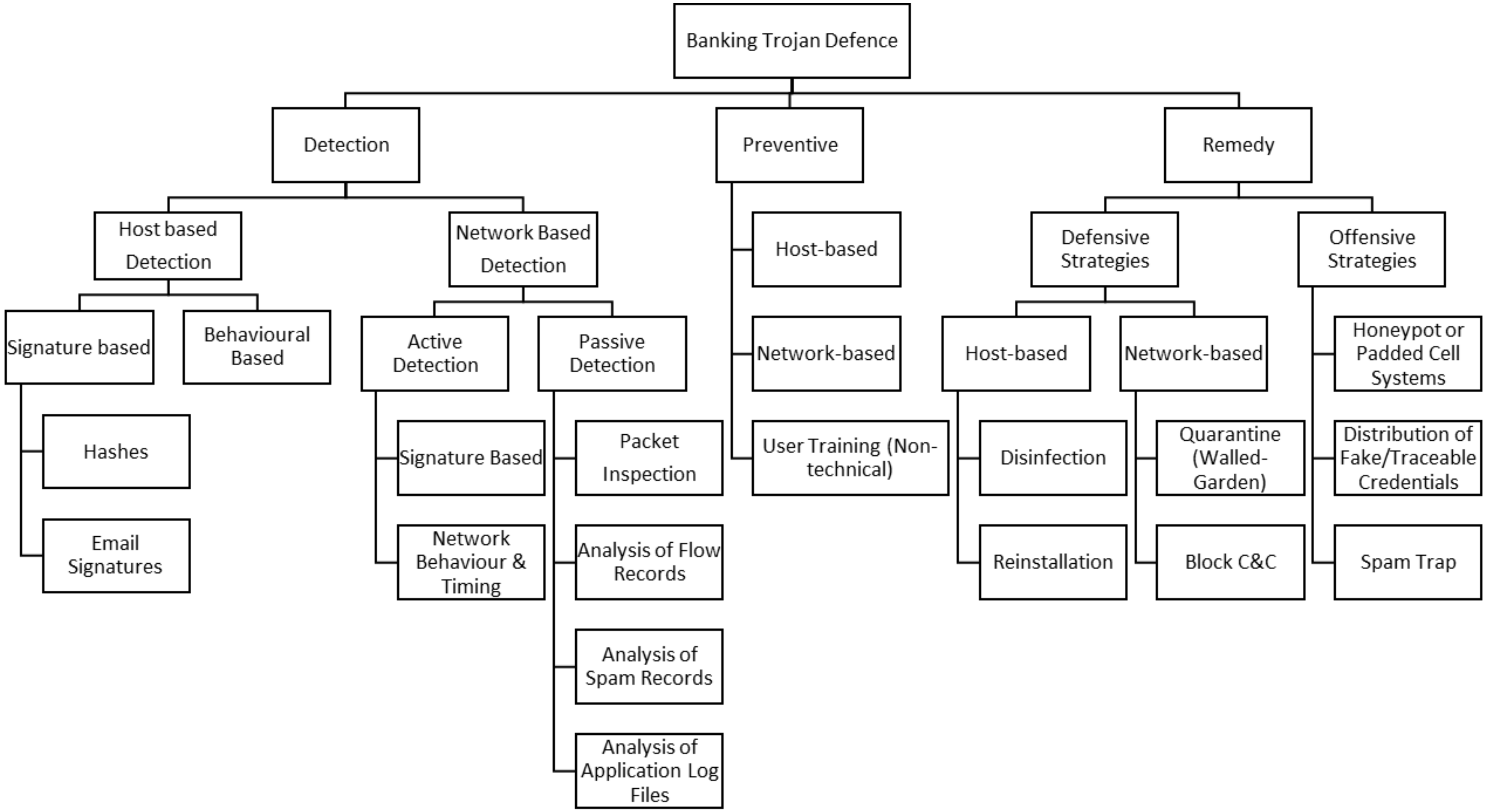

**Figure 4.** Banking Trojan Defence Taxonomy.

# 1. Detection

Detection techniques provide opportunities to identify a malicious attack in its early stages i.e. during reconnaissance, weaponization or delivery [20], [31]. Detection can be conducted on host or network level.

## 1.1. Host Based Detection

The most common host based detection tools are Antivirus (AV) software which utilise two detection methods namely signatures-based or behavioural-based [2], [48], [55].

***Signature-Based; S***ignature can be list of domains which are known to be malicious or hashes of the known malware [48]. These signatures are stored in a large repository for the sake of comparison for malware detection [28], [48], [88].

***Hashes;*** hashes can either be MD5 or SHA, this mechanism is mostly preferable due to it consistence and accuracy [48]. AV vendors usually utilise a generic signature which is hash of the code segment to detect malware samples belong to the same family [48], [49].

***Email Signatures;*** Trojans can spoof sender's address to reflect well known organisations such as banks, power companies and post office. A signature based on these known malicious trends can be created along with a blocking rule on an IDS to prevent users from receiving such emails [111], [112]. Although this strategy can be easily bypassed by adversaries, it is still effective in the absence of adversaries' knowledge of being shunned.

***Behavioural Based;*** this type of detection aims to identify actions performed by malwares and trends of events rather than syntactic markers to identify an attack [38], [88], [113]. Behavioural/heuristic technique of detection can be either static or automated, although automated techniques are more practical than static ones [114]. While we appreciate the applicability of signature based detection in real life application, heuristic becomes more prominent as they can provide deep and effective analysis to determine the malicious behaviours of activities [114]–[118]. The reason this technique is most effective is because the behaviour of each Trojan can be emulated in a sandbox and analysed. Depending on methodology used to reduce false positives [117], [118], the technique may often be associated with the disadvantage and consumption of resources. With the rise in Internet of Things (IoT) behavioural detection can be enhanced with the intel gathered by this feature and that of machine learning capabilities [117]–[119].

## 1.2. Network Based Detection

Network based detection techniques can be categorised into active or passive [2], [81].

***Active Detection;*** this technique involves interaction with information resources. Once a malware is detected a comparison is made with a set of previously generated malicious signatures [120]. The main active detection techniques are signature-based or network behaviour and timing based.

***Signature-Based:*** IDS like SNORT uses sets of signatures to determine if an event is malicious or not [81]. The comparison could be based on simple features such as header of an email address, the sender's domain name or more complicated such as comparing an email with the blacklisted domains or known malicious activities. Of course this technique is vulnerable to zero-day attacks [121] as signatures will not be available in the database.

***Network Behaviour & Timing;*** in this technique, the comparison is based on features such as if the connection is coming from a known malicious IP address, known vulnerable ports, or even analysis of dishonest requests. Moreover, timing of sent requests can be used as a feature to determine if they are robot made or human made i.e. multiple request in very short period of time usually signifies a bot origin [122]–[124].

***Passive Detection;*** in this technique, we achieve detection through monitoring activities initiated by applications to distinguish between malicious Trojans and benign apps. In passive detection, the defender remain undetected and gather as much information as possible without spooking the adversary[120]. Following techniques can be used to hat detect malicious activities in C&C and Data exfiltration stage of an attack:

***Packet Inspection;*** when a Trojan is in rallying or exfiltration phase many indications of maliciousness can be observed by finding connections to a known blacklisted domains or IP addresses [125]. We can also observe payloads of individual packets from contents of GET or POST requests to deduce what the Trojan is sending or receiving [33].

***Analysis of Flow Records;*** this technique considers the flow of the entire traffic instead of individual packets. In this technique several attributes of communication streams are looked at such as source and destination addresses, port numbers, communication protocol, the duration of a session, and the cumulative size and number of transmitted packets[125]. With such attributes a detection matrix can be constructed to detect malicious activities not only in rallying and data exfiltration stage of the banking Trojans but also in the delivery stage.

***Analysis Of Spam Records;*** As Trojans are mostly delivered though unsolicited emails (referred to as spam) [125], analysis of these emails can lead to establishment of a pattern or a detection matrix . Filters can be set to scan for specific fingerprints from subject headers, message bodies, white and even black lists email addresses [111], [112].

***Analysis of Application Log Files;*** analysis of logs is not so different than spam emails, except that logs come from different sources i.e. firewalls, applications, servers etc.! The biggest challenge is analysing logs on a timely manner to detect malicious or just abusive behaviour of an application or users of the systems generating the logs [126], [127]. When a Trojan run on a machine one can discover from a log file what files the Trojan has created, deleted, or DLLs that are injected etc.

## 2. *Preventive*

After a successful detection and information gathering of any attack, what follows next is putting in place safeguards to prevent future attacks. Preventive techniques can be categorized into host-based, network-based and user training.

### 2.1. Host Based

When considering prevention at the level of host machine, anti-malware, anti-spyware and other tools are the most commonly used [48]. With good enough detection signatures this technique is very useful though not all Trojans can be detected and prevented easily due to the variant created by adversaries. Recent Symantec security product [85] has highlighted its capability in detecting and preventing attacks from Dridex malware. The product has highlighted the capability of scanning attachments of the incoming emails on host level and preventing users from opening malicious attachments.

Furthermore, extensive access controls on confidential files should be considered to prevent Trojans sniffing around systems to gain usable intel. An Attribute-Based Encryption (ABE) which allows user to define access to a specific piece of data depending on either a true or false outcome of a boolean expression on sets of attributed defined in a policy [128]. An ABE has been suggested by researcher as amongst powerful cryptographic tool that allows fine grained access control on data that can prevent malware from stealing your data [128]. On top of tools like ABE, other strong data encryption at rest ciphers can be useful to support host based preventive measures.

### 2.2. Network Based

When casting a wider net for prevention on your network, things tends to get complex but in a long run it’s much easier managing attacks from their point of entry to your network rather than managing attacks on individual host machines. Some organisations have achieved this through implementation of SMTP and HTTP/SSL proxies [49], whereby all traffic from the internet towards organisations' internal network must pass these gateways which provides extra layer of security. Moreover, these gateways can be configured with email filtering rules i.e. any incoming email will be assessed in correlation with the rules set and the system will decide if an email should be allowed into the network or not. Some organisations have done this by deploying network Intrusion Prevention Systems (IPS). While one may argue the importance of having IPS while firewall already exist, the truth is firewalls themselves have limitations. Firewalls can block traffics basing on IP blocks and ports while an IPS monitors all traffic entering and exiting the network [129]. With organisations hosting webserver applications like Apache and Microsoft IIS, relying on firewall alone is ill-advised hence, having an IPS on your layered defensive security provides assurance while maintain functionality. The advantage of using an IPS as opposed to IDS is the ability of IPS to stop or redirect attacks pre-emptively [129]. With this feature an IPS is capable of redirecting attacks to the honeypot machine luring the attackers to believe success while giving the analyst ample time to analyse the attack and understand it in depth.

### 2.3. User Training

With ISO 27001 standard in play as a minimum requirement the provision of awareness training to all employees is mandatory. If users understand that most of the current Trojan attacks are based on spam campaign and receive appropriate training and act accordingly many Trojan attacks were never happened [130]. Furthermore, understanding that a Microsoft document or attachment may contain embedded malware will also give users a perspective on how to treat received attachments from a known or unknown sender.

## 3. *Remedial*

As part of defence strategies, a consideration of safeguard failure should always be accounted. In the event safeguards against Trojan attack were not successful, we need sets of remedial actions that enable us to revert the system to a point of proper functionality. Remedial activities can be divided into defensive and offensive. By defensive we are looking at the actions to be taken after infection to bring your system back to normal, prevent spreading of the infection and repetition of the infection. On offensive side, we are looking at the actions that allow us to gather more intelligence about an attack, get an insight of the attacker and possibly trace the attack to its origin with the sole purpose of dismantling the Trojan infrastructure.

### 3.1. Defensive Strategies

Once an infection has been identified, we need strategies to bring the system back to its normal functionalities. Even these steps have merits and demerits but with a good plan we can recover from an attack as quickly as possible. We further classify defensive strategies into those dealing with an individual host machine and those concerned with the entire network.

***Host-based:*** on the host machine, several steps can be taken to remediate from a Trojan infection among which disinfection or reinstallation can be considered.

***Disinfection;*** this simply means removing the Trojan from the infected machine. We know the ultimate non-functionality of Trojan is receiving a STOP or KILL command from its controller. If by technical means an investigator can figure out the Trojan self-destruction mechanism, it can be triggered to remove the infection [125]. However, there are off-shelf programs such as Trojan removal kits that can be used although their ultimate effectiveness is hard to guarantee.

***Reinstallation;*** A complete reinstallation of an infected machine should be considered as the last option if disinfection is not viable as it takes longer time and involves much more activities. If a disk image of a clean installation is available, the process can be well shortened, otherwise a clean installation from preferred OS disk should be considered.

***Network-based:*** when the infection has spread to more than just one machine then the strategy should also span out accordingly. There are two main techniques for network based remedy of Trojans as follows:

***Quarantine (Walled-garden):*** isolating machines with symptoms of malware infection either by seeing their connections to a known C&C or malicious downloads and uploads [36], [125] is referred as walled-garden. Isolated machines are denied access to any other website except for the ones for remedial activates until the machine(s) is/are properly cleaned and pass security policies checks.

***Block C&C:*** Trojans are receiving instructions from a controller (C&C server) and send collected information to C&C. Once a C&C server IP address is identified we need to block communication to the address to prevent Trojans from receiving instructions or sending information to the controller. The other way to deal with this is by only allowing traffic to known addresses and block all other communications to unknown addresses in the network.

## 3.2. Offensive strategies

The idea here is intelligent gathering while protecting live environment, and ultimately paralysing the Trojan botnet. With the utilisation of virtual machines or cloud based simulation platforms security experts can gather sufficient evidence up to successful prosecution provided the integrity of the evidence is accounted for [6], [70]. With offensive strategies skiing on the edge of the line of legality and ethics, one needs to be very careful as attacking back is not only unethical but illegal too.

***Honeypot Or Padded Cell Systems:*** Honeypots are legal traps, intentionally deployed in a network to detect or deflect unauthorised access to a system [2], [125], [131]. For clarity, this is not a legal advice for legal advice please consult your legal counsel/lawyer. Honeypots are often associated with three concepts when it comes to legal conception i.e. enticement, entrapment, and privacy. From legal litigation the three are applicable only if the aim was to prosecute someone, other than that honeypots are legal and can be used [132]. Honeypots are useful tools in understanding attacker’s intentions,

Honeypots contains no business data so any access to the honeypot is considered malicious. Honeypots are used to research and study attacks and discover new information about the strategies and practices used by malware creators. With honeypots two kinds of information can be gathered. First is type of attack vectors in operating systems and software used along with the actual exploit code and second are attackers activities performed on an exploited machine so as to come up with better defence tactics for such attacks [125]. Also institutional honeypots can be used for developing better training programs to enhance security awareness [133].

Padded cells and honeypot use similar concept, but the distinct difference between the two systems is that, padded cells are highly protected and cannot be easily compromised as compared to honeypots [134]. In other words, padded cell is a hardened honeypot and operates in tandem with traditional IDS. With such systems in place, Trojan understanding becomes clearer to an organisation security team and better defence mechanisms are constructed with high percentage of certainty about future trends.

***Distribution of Fake Credentials;*** The main goal of a banking Trojans is online banking credentials, or credit and debit card details [125]. By identifying where stolen information is being submitted by a Trojan (also known as drop zone), crafted false and target-oriented information can be injected into these drop zones [36], [125], [135]. Therefore, mistrust between botmaster and their customer will be created and possibly run some players out of business and the entire malicious infrastructure may become useless. By associating the credentials with a monitored account, banks can trace the money and find where it is being transferred to and perhaps apprehend a botmaster [125].

***Spam Trap;*** Most Trojans are being distributed by spam campaigns [125]. Spam trap can be used to capture as much of the spam emails for further analysis. Spam trap is considered as the email address with no functionality other than receiving unsolicited emails [136], [137]. The simplest setup of this trap is by advertising an email into many newsletters and forum to make it known so attackers will send their spam to that address. The evaluation of spam message along with its attachment and included links can assist security analysts to deduce information about malware families and even lead to detection of unknown families.

Most of the defence techniques like IDS and/or IPS are used both for detection and prevention but we should understand their limitations too. Defence techniques highly depend on correct and up to date information. For instance, detecting an embedded Trojan in an office document by an IDS system can be difficult but if the IDS system is properly configured on both network perimeter and host the chance of detection increases. Likewise, detection of signatures highly depends on the software in use, but with the consideration of up to date signatures and continuous maintenance/update of the software, organisation may not face difficulties in detecting attacks.

## V. Concluding Remarks

While evolutionary computational intelligence approaches are a viable approach to designing intelligent and effective malware detection and mitigation solutions, having a features taxonomy can help inform the design of such approaches by reducing the impreciseness, subjectivity, and knowledge uncertainty in decision making process.

In this paper, we presented the first Cyber Kill Chain (CKC)-based taxonomy which details banking Trojans features, and validated the taxonomy based on real-world samples. Cyber defenders can then use the taxonomy to design their banking Trojans detection and mitigation strategy, including those based on evolutionary computational intelligence approaches. It is notable, that since as our proposed taxonomies are based on malware observed in financial sector in the UK, extra cautions should be taken in extending our results to other contexts.

In the future, we plan to extend this taxonomy to cover other malware families as well as implementing the defence taxonomy using evolutionary computational intelligence (i.e. proof-of-concept) for deployment and evaluation. Moreover, building similar taxonomies for forensics investigation, incident handling and threat hunting of Banking Trojans are other interesting future works of this research.

### Acknowledgment

The views and opinions expressed in this article are those of authors alone and not the organizations with whom authors are or have been associated or supported. We thank VirusTotal for graciously providing us with a private API key to access their data to prepare our dataset. This work is partially supported by the European Council International Incoming Fellowship (FP7-PEOPLE-2013-IIF) grant.

**Appendix I**

| Allocated Name | MD5 Hash | Allocated Name | MD5 Hash | Allocated Name | MD5 Hash |
|---|---|---|---|---|---|
| Trojan1 | cd238c1dab76a4336db727cdcbdcfc13 | Trojan44 | cacb79e05cf54490a7067aa1544083fa | Trojan87 | f5aee45ce06f6d9f9210ae28545a14c6 |
| Trojan2 | c8247ee42c27364c3c33def68102241e | Trojan45 | 265f3b610aed3745ba19fd795a748e57 | Trojan88 | b8d83b04a06b6853ad3e79a977dd17af |
| Trojan3 | 2feb35e572e0339735804c42184f422b | Trojan46 | 59b8bdd04ca78f3ac74f0d2bd414a9cf | Trojan89 | 6e5654da58c03df6808466f0197207ed |
| Trojan4 | eadafc9b1891c74bc13e09c46c4a40c9 | Trojan47 | c9678ee6a19547fb213126fba9f6036d | Trojan90 | 1fac282d89e9af6fd548db2c71124c38 |
| Trojan5 | c96abe929eb587cb2913b638df368b3b | Trojan48 | eca90bf0af7db8ac5ec7993761f97f49 | Trojan91 | 5bddf5271b1472eca61a6a2d66280020 |
| Trojan6 | 01078f660f979b30e4624e57cf986b6c | Trojan49 | 514622A1797D9916637EED7833CFD5AF | Trojan92 | eb7df68bd7eb7cf2968cf541af3472d6 |
| Trojan7 | 104528e67f01168e12cdac550fc43260 | Trojan50 | fc4a4449246bbec022339618665a1976 | Trojan93 | 50e3407557500fcd0d81bb6e3b026404 |
| Trojan8 | 000b2347b8fe2cc625861d7ed3ec834f | Trojan51 | f3b124852d90b5b32d03131e77b3ac2c | Trojan94 | c6cefd2923164aa14a3bbaf0dfbea669 |
| Trojan9 | dc92858693f62add2eb4696abce11d62 | Trojan52 | fbca3b9e23ef33b25e74be7511dcecc1 | Trojan95 | b227c91fbc1ba56e9f01ab4f1e2e502f |
| Trojan10 | 96f3aa2402daf9093ef0b47943361231 | Trojan53 | 585381110056b63957a22e6aef59a31e | Trojan96 | 2845499946fd5882f94cc9a4375b364a |
| Trojan11 | e4cc002a95caaf4481cb7140bbe96c58 | Trojan54 | 66120bf739f2d53ef930194165eb5d09 | Trojan97 | 37ceca4ac82d0ade9bac811217590ecd |
| Trojan12 | a4e14c88da9e1a74cd7c26ded99b6a0a | Trojan55 | 66120bf739f2d53ef930194165eb5d09 | Trojan98 | D752837D0EE0E5D49D1F72F52279948E |
| Trojan13 | e8cd8be37e30c9ad869136534f358fc5 | Trojan56 | 0D02257EC18B92B3C1CF58B8CB6B3D37 | Trojan99 | 289af95f99f58c751a7d1d0a26d7cdb3 |
| Trojan14 | 3e3a09644170ad3184facb4cace14f8a | Trojan57 | 48d496afc9c2c123e1ab0c72822a7975 | Trojan100 | 289af95f99f58c751a7d1d0a26d7cdb3 |
| Trojan15 | 03ab12e578664290fa17a1a95abd71c4 | Trojan58 | 373c9e5461c2b234f70e4d6102198eff | Trojan101 | e25a05d3fecceb14667048c07494d65f |
| Trojan16 | e46dcc4a49547b547f357a948337b929 | Trojan59 | b5d68075a093c263fb3392cb92d92bef | Trojan102 | 4c3d0e0a944fc6755a28452e913e0347 |
| Trojan17 | 2ecf5e35d681521997e293513144fd80 | Trojan60 | ea4bbf027eb58b92566eb4d98002f976 | Trojan103 | 1de3889fde95e695adf6eadcb4829c6d |
| Trojan18 | 6c784bec892ce3ef849b1f34667dccac | Trojan61 | e14f089df621262bcbf172b5a5346d33 | Trojan104 | e4bb8a66855f6987822f5aca86060f2c |
| Trojan19 | 673626be5ea81360f526a378355e3431 | Trojan62 | 148112df459ba40b9127f7d4f1c08df2 | Trojan105 | 7f0076993f2d8a4629ea7b0df5b9bddd |
| Trojan20 | 02492b954b48f13412a844d689d064f1 | Trojan63 | ee265704cf0a371029ed28e958e06549 | Trojan106 | 39837c6ba74a922f935d184d8f0f0d7b |
| Trojan21 | 1FC2ABEC9C754E8CC1726BF40E0B3533 | Trojan64 | 154faec2f2ac9c0fb028680e0e0ee78c | Trojan107 | 999ebe8b65caf99faf1172074d7e5d3c |
| Trojan22 | e52a8d15ee08d7f8b4efca1b16daaefb | Trojan65 | a49149e8822c5c692cd71736a513d268 | Trojan108 | 74dc37b7aabf745eac1d5fc65428488e |
| Trojan23 | e52a8d15ee08d7f8b4efca1b16daaefb | Trojan66 | 53ba28120a193e53fa09b057cc1cbfa2 | Trojan109 | a5c52bd47f7fdfd54a2584a669eabe59 |
| Trojan24 | DA26ED1B6FE69D15A400B3BC70001918 | Trojan67 | a29122dfa93bcac56ab9e5e05ac1d41a | Trojan110 | 6c14578c2b77b1917b3dee9da6efcd56 |
| Trojan25 | 6aa26f04b22b284dda148ce317f53de8 | Trojan68 | 20343AE1698C45FB3ECE073745D28D4C | Trojan111 | FDD95B4CC10B536934486C7D3FDEE04F |

| Allocated Name | MD5 Hash | Allocated Name | MD5 Hash | Allocated Name | MD5 Hash |
|---|---|---|---|---|---|
| Trojan26 | e1c7eccc8fec00a10c1e0cd65e443635 | Trojan69 | eb19dfe2116be14283c254a16a786482 | Trojan112 | 5ab2a67268b3362802a13594edafbd2e |
| Trojan27 | 0864bc6951795b86d435176c3320a8bc | Trojan70 | ace2a3e0ca6bdfa6331a6e7d519ab1e7 | Trojan113 | 32a34ce536ca62c61cc05ce3e3f3c54f |
| Trojan28 | 0864bc6951795b86d435176c3320a8bc | Trojan71 | ace2a3e0ca6bdfa6331a6e7d519ab1e7 | Trojan114 | aab74722020e631147836fc009f9419d |
| Trojan29 | ab40142988527fe6ce585a9fdfce56ca | Trojan72 | dcc7f58bff80b337e5e7723b2ac9dad7 | Trojan115 | 264E49C78F3693F4DEEBA9D62F3F5C89 |
| Trojan30 | a68b72fbfb76964261a3601daa270647 | Trojan73 | d41421a918ce05632374081c33879d4c | Trojan116 | 4ba35d78df77a4d5ad1207cdeeef78b3 |
| Trojan31 | 23964bc22c2c81f9a41fb9f747a6c995 | Trojan74 | 461689d449c7b5a905c8404d3a464088 | Trojan117 | c66dfa4304d9782f19cab27379191f7a |
| Trojan32 | fd7b410fd7936dd51c4b72ef4047c639 | Trojan75 | 02cfa3e6fdb4301528e5152de76b2abf | Trojan118 | cfeab92b4e304d188c3e6f81d6d6925b |
| Trojan33 | 0316dbd20fbfd5a098cd8af384ca950f | Trojan76 | 107a3bef0da9ab2b42e3e0f9f843093b | Trojan119 | 741fad6dabdc81f485c6fbd8a8ce125d |
| Trojan34 | 6e8f48e7d53ac2c8f7b863078e9050b2 | Trojan77 | 4c7f72fc16ac8daf5237cfc4e5546ac0 | Trojan120 | dd93f9f9d2ec75096ed843e386d68f4c |
| Trojan35 | FC1E5521A5F2479EA3226288B6205300 | Trojan78 | 107a3bef0da9ab2b42e3e0f9f843093b | Trojan121 | a8a42968a9bb21bd030416c32cd34635 |
| Trojan36 | dcb019624fb8e92eb26adf2bef77d46c | Trojan79 | 716d1dc7285b017c2dbc146dbb2e319c | Trojan122 | 6bd532a798f5b473e4237342c3d4d580 |
| Trojan37 | 8b288305733214f8e0d95386d886af2d | Trojan80 | 412ce577521a560459cd711f5966caf4 | Trojan123 | b4fb40b3dfa5780732d599eba6023309 |
| Trojan38 | f9c00d3db5fa6cd33bc3cd5a08766ad0 | Trojan81 | 818231cb0be9bf597d33013edb85e1a7 | Trojan124 | f71529ae0cab12fa089b91e333ac5d6f |
| Trojan39 | 1fbf5be463ce094a6f7ad345612ec1e7 | Trojan82 | dd7adc5b140835dc22f6c95694f9c015 | Trojan125 | c97fcb5f276542ac719fef3d32fbd2bf |
| Trojan40 | D73D599EF434D7EDAD4697543A3E8A2B | Trojan83 | 3fcc933847779784ece1c1f8ca0cb8e4 | Trojan126 | f23c05c44949c6c8b05ab54fbd9cee40 |
| Trojan41 | d5e717617400b3c479228fa756277be1 | Trojan84 | d2f825ecfb3d979950b9de92cbe29286 | Trojan127 | 75b6411071a27959394ffba9ecdea4a7 |
| Trojan42 | d5e717617400b3c479228fa756277be1 | Trojan85 | af15ba558c07f8036612692122992aad | | |
| Trojan43 | 6932a004ce3ad1ad5ea30f43a31b0285 | Trojan86 | 7008675da5c1b0a6b59834d125fafa45 | | |

## Appendix II

| Trojan Al-located Name | Banking Trojans Features | | | | | | | | | | |
|---|---|---|---|---|---|---|---|---|---|---|---|
| | Weaponization | | | Delivery | | Exploitation | Installation | | C&C | | Data Exfil-tration |
| | Host-Based Evasion | Anti-forensics | | Email Attachments | | | | | | | |
| | Rootkit | Dead-Code | Utilising Packers | Macros | PDF(JS) | API Hooking | Heap Spraying | DLL Side loading | Hardcoded IP | Domain Name | Backdoors |
| Trojan1 | ✓ | | | | ✓ | | | | ✓ | | ✓ |
| Trojan2 | ✓ | | | ✓ | | ✓ | | | ✓ | | ✓ |
| Trojan3 | ✓ | | | ✓ | | | | | ✓ | | ✓ |
| Trojan4 | ✓ | | | ✓ | | ✓ | | | ✓ | | ✓ |
| Trojan5 | ✓ | | | ✓ | | ✓ | | | ✓ | | ✓ |
| Trojan6 | | | | | ✓ | | ✓ | | ✓ | | ✓ |
| Trojan7 | ✓ | | | | ✓ | | ✓ | | ✓ | | ✓ |
| Trojan8 | ✓ | | | | ✓ | ✓ | ✓ | | ✓ | | ✓ |
| Trojan9 | | | | ✓ | | | | | ✓ | | |
| Trojan10 | | | ✓ | ✓ | | ✓ | | ✓ | ✓ | ✓ | ✓ |
| Trojan11 | | | ✓ | ✓ | | ✓ | | ✓ | ✓ | ✓ | ✓ |
| Trojan12 | | | ✓ | ✓ | | ✓ | | ✓ | ✓ | ✓ | ✓ |
| Trojan13 | | | ✓ | ✓ | | ✓ | | ✓ | ✓ | ✓ | ✓ |
| Trojan14 | | | ✓ | ✓ | | ✓ | | ✓ | ✓ | ✓ | ✓ |
| Trojan15 | | | ✓ | ✓ | | ✓ | | ✓ | ✓ | ✓ | ✓ |
| Trojan16 | | | ✓ | ✓ | | ✓ | | ✓ | ✓ | ✓ | ✓ |
| Trojan17 | | | ✓ | ✓ | | ✓ | | ✓ | ✓ | ✓ | ✓ |
| Trojan18 | | | ✓ | ✓ | | ✓ | | ✓ | ✓ | ✓ | ✓ |
| Trojan19 | | | ✓ | ✓ | | ✓ | | ✓ | ✓ | ✓ | ✓ |
| Trojan20 | | | ✓ | ✓ | | ✓ | | ✓ | ✓ | ✓ | ✓ |
| Trojan21 | | | ✓ | ✓ | | ✓ | | ✓ | ✓ | ✓ | ✓ |
| Trojan22 | | | ✓ | ✓ | | ✓ | | ✓ | ✓ | ✓ | ✓ |
| Trojan23 | | | ✓ | ✓ | | ✓ | | ✓ | ✓ | ✓ | ✓ |
| Trojan24 | | | ✓ | ✓ | | ✓ | | ✓ | ✓ | ✓ | ✓ |
| Trojan25 | | | ✓ | ✓ | | ✓ | | ✓ | ✓ | ✓ | ✓ |
| Trojan26 | | | | ✓ | | ✓ | | ✓ | ✓ | ✓ | ✓ |
| Trojan27 | | | ✓ | ✓ | | ✓ | | ✓ | ✓ | ✓ | ✓ |
| Trojan28 | | | ✓ | ✓ | | ✓ | | ✓ | | ✓ | ✓ |
| Trojan29 | | | | ✓ | | ✓ | ✓ | | | ✓ | |
| Trojan30 | | | | ✓ | | ✓ | ✓ | | | ✓ | |
| Trojan31 | | | | ✓ | | ✓ | ✓ | | | ✓ | |
| Trojan32 | | | | ✓ | | ✓ | | ✓ | | ✓ | ✓ |
| Trojan33 | | | | ✓ | | ✓ | | ✓ | | ✓ | ✓ |
| Trojan34 | | ✓ | ✓ | ✓ | | | | | ✓ | ✓ | ✓ |
| Trojan35 | | ✓ | ✓ | ✓ | | | | | ✓ | ✓ | ✓ |
| Trojan36 | | | | ✓ | | ✓ | | ✓ | | ✓ | ✓ |
| Trojan37 | | | | ✓ | | ✓ | | ✓ | | ✓ | ✓ |

| Trojan Allocated Name | Banking Trojans Features | | | | | | | | | | |
|---|---|---|---|---|---|---|---|---|---|---|---|
| | Weaponization | | | Delivery | | Exploitation | Installation | | C&C | | Data Exfiltration |
| | Host-Based Evasion | Anti-forensics | | Email Attachments | | | | | | | |
| | Rootkit | Dead-Code | Utilising Packers | Macros | PDF(JS) | API Hooking | Heap Spraying | DLL Side loading | Hardcoded IP | Domain Name | Backdoors |
| Trojan38 | | ✓ | ✓ | ✓ | | | | | ✓ | ✓ | ✓ |
| Trojan39 | | ✓ | ✓ | ✓ | | ✓ | ✓ | | | ✓ | |
| Trojan40 | | ✓ | ✓ | ✓ | | ✓ | | ✓ | | ✓ | ✓ |
| Trojan41 | | | ✓ | ✓ | | | | | | ✓ | ✓ |
| Trojan42 | | | ✓ | ✓ | | | | | | ✓ | ✓ |
| Trojan43 | | | ✓ | ✓ | | | | ✓ | ✓ | ✓ | ✓ |
| Trojan44 | | | ✓ | ✓ | | | | ✓ | ✓ | ✓ | ✓ |
| Trojan45 | | | ✓ | ✓ | | | | | ✓ | | ✓ |
| Trojan46 | ✓ | | ✓ | ✓ | | | | | ✓ | ✓ | |
| Trojan47 | | | ✓ | ✓ | | | | | ✓ | ✓ | ✓ |
| Trojan48 | ✓ | ✓ | ✓ | ✓ | | ✓ | | | ✓ | | ✓ |
| Trojan49 | ✓ | ✓ | ✓ | ✓ | | ✓ | | | ✓ | | ✓ |
| Trojan50 | ✓ | ✓ | ✓ | ✓ | | ✓ | | | ✓ | ✓ | ✓ |
| Trojan51 | | ✓ | ✓ | ✓ | | ✓ | | | ✓ | ✓ | ✓ |
| Trojan52 | | ✓ | ✓ | ✓ | | ✓ | | | ✓ | ✓ | ✓ |
| Trojan53 | | | | ✓ | | | | | | | ✓ |
| Trojan54 | | ✓ | ✓ | ✓ | | ✓ | | | ✓ | | ✓ |
| Trojan55 | | ✓ | ✓ | ✓ | | ✓ | | | ✓ | | ✓ |
| Trojan56 | | ✓ | ✓ | ✓ | | | | | | ✓ | ✓ |
| Trojan57 | ✓ | ✓ | ✓ | | ✓ | ✓ | ✓ | | ✓ | ✓ | ✓ |
| Trojan58 | | ✓ | | ✓ | | ✓ | | | ✓ | ✓ | ✓ |
| Trojan59 | ✓ | ✓ | ✓ | ✓ | | ✓ | | | ✓ | ✓ | ✓ |
| Trojan60 | ✓ | ✓ | ✓ | ✓ | | ✓ | | | ✓ | ✓ | ✓ |
| Trojan61 | ✓ | ✓ | | ✓ | | ✓ | | | ✓ | ✓ | ✓ |
| Trojan62 | ✓ | ✓ | | ✓ | | ✓ | | | ✓ | ✓ | ✓ |
| Trojan63 | | | | ✓ | | ✓ | | | ✓ | | |
| Trojan64 | | | | ✓ | | | | | | | |
| Trojan65 | ✓ | ✓ | ✓ | ✓ | | ✓ | | | ✓ | ✓ | ✓ |
| Trojan66 | ✓ | ✓ | ✓ | ✓ | | ✓ | | | ✓ | ✓ | ✓ |
| Trojan67 | ✓ | ✓ | ✓ | ✓ | | ✓ | | | ✓ | ✓ | ✓ |
| Trojan68 | ✓ | ✓ | ✓ | ✓ | | ✓ | | | ✓ | ✓ | ✓ |
| Trojan69 | ✓ | ✓ | ✓ | ✓ | | ✓ | | | ✓ | ✓ | ✓ |
| Trojan70 | ✓ | ✓ | ✓ | ✓ | | ✓ | | | ✓ | ✓ | ✓ |
| Trojan71 | ✓ | ✓ | ✓ | ✓ | | ✓ | | | ✓ | ✓ | ✓ |
| Trojan72 | ✓ | | ✓ | ✓ | | ✓ | | | ✓ | ✓ | ✓ |

| Trojan Al-located Name | Banking Trojans Features | | | | | | | | | | |
|---|---|---|---|---|---|---|---|---|---|---|---|
| | Weaponization | | | Delivery | | Exploitation | Installation | | C&C | | Data Exfil-tration |
| | Host-Based Evasion | Anti-forensics | | Email Attachments | | | | | | | |
| | Rootkit | Dead-Code | Utilising Packers | Macros | PDF(JS) | API Hooking | Heap Spraying | DLL Side loading | Hardcoded IP | Domain Name | Backdoors |
| Trojan73 | | | ✓ | ✓ | | ✓ | | | ✓ | | |
| Trojan74 | | | ✓ | ✓ | | ✓ | | | ✓ | | |
| Trojan75 | | | ✓ | ✓ | | ✓ | | | ✓ | | |
| Trojan76 | ✓ | ✓ | ✓ | ✓ | | ✓ | | | ✓ | ✓ | ✓ |
| Trojan77 | ✓ | ✓ | ✓ | ✓ | | ✓ | | | ✓ | ✓ | ✓ |
| Trojan78 | ✓ | ✓ | ✓ | ✓ | | ✓ | | | ✓ | ✓ | ✓ |
| Trojan79 | ✓ | ✓ | ✓ | ✓ | | ✓ | | | ✓ | ✓ | ✓ |
| Trojan80 | ✓ | ✓ | ✓ | ✓ | | ✓ | | | ✓ | ✓ | ✓ |
| Trojan81 | ✓ | ✓ | ✓ | ✓ | | ✓ | | | ✓ | ✓ | ✓ |
| Trojan82 | ✓ | ✓ | ✓ | ✓ | | ✓ | | | ✓ | ✓ | ✓ |
| Trojan83 | ✓ | ✓ | ✓ | ✓ | | ✓ | | | ✓ | ✓ | ✓ |
| Trojan84 | ✓ | ✓ | ✓ | ✓ | | ✓ | | | ✓ | ✓ | ✓ |
| Trojan85 | ✓ | ✓ | ✓ | ✓ | | ✓ | | | ✓ | ✓ | ✓ |
| Trojan86 | ✓ | ✓ | ✓ | ✓ | | ✓ | | | ✓ | ✓ | ✓ |
| Trojan87 | ✓ | ✓ | ✓ | ✓ | | ✓ | | | ✓ | ✓ | ✓ |
| Trojan88 | ✓ | ✓ | ✓ | ✓ | | ✓ | | | ✓ | ✓ | ✓ |
| Trojan89 | ✓ | ✓ | ✓ | ✓ | | ✓ | | | ✓ | | ✓ |
| Trojan90 | ✓ | ✓ | ✓ | ✓ | | ✓ | | | ✓ | | ✓ |
| Trojan91 | ✓ | ✓ | ✓ | ✓ | | ✓ | | | ✓ | ✓ | ✓ |
| Trojan92 | ✓ | ✓ | ✓ | ✓ | | ✓ | | | ✓ | ✓ | ✓ |
| Trojan93 | ✓ | ✓ | ✓ | ✓ | | ✓ | | | ✓ | ✓ | ✓ |
| Trojan94 | ✓ | ✓ | ✓ | ✓ | | ✓ | | | ✓ | ✓ | ✓ |
| Trojan95 | ✓ | ✓ | ✓ | ✓ | | ✓ | | | ✓ | ✓ | ✓ |
| Trojan96 | ✓ | ✓ | ✓ | ✓ | | ✓ | | | ✓ | ✓ | ✓ |
| Trojan97 | ✓ | ✓ | ✓ | ✓ | | ✓ | | | ✓ | ✓ | ✓ |
| Trojan98 | ✓ | ✓ | ✓ | ✓ | | ✓ | | ✓ | ✓ | | |
| Trojan99 | ✓ | ✓ | ✓ | ✓ | | ✓ | | | ✓ | ✓ | ✓ |
| Trojan100 | ✓ | ✓ | ✓ | ✓ | | ✓ | | | ✓ | ✓ | ✓ |
| Trojan101 | ✓ | ✓ | ✓ | ✓ | | ✓ | | | ✓ | ✓ | ✓ |
| Trojan102 | ✓ | ✓ | ✓ | ✓ | | ✓ | | ✓ | ✓ | | |
| Trojan103 | ✓ | ✓ | ✓ | ✓ | | ✓ | | | ✓ | ✓ | ✓ |
| Trojan104 | | | | ✓ | | | | | ✓ | | |
| Trojan105 | ✓ | ✓ | ✓ | ✓ | | ✓ | | | ✓ | ✓ | ✓ |
| Trojan106 | ✓ | ✓ | ✓ | ✓ | | ✓ | | | ✓ | ✓ | ✓ |

| Trojan Allocated Name | Banking Trojans Features | | | | | | | | | | |
|---|---|---|---|---|---|---|---|---|---|---|---|
| | Weaponization | | | Delivery | | Exploitation | Installation | | C&C | | Data Exfiltration |
| | Host-Based Evasion | Anti-forensics | | Email Attachments | | | | | | | |
| | Rootkit | Dead-Code | Utilising Packers | Macros | PDF(JS) | API Hooking | Heap Spraying | DLL Side loading | Hardcoded IP | Domain Name | Backdoors |
| Trojan107 | ✓ | ✓ | ✓ | ✓ | | ✓ | | | ✓ | ✓ | ✓ |
| Trojan108 | | | | ✓ | | | | | ✓ | | |
| Trojan109 | ✓ | ✓ | ✓ | ✓ | | ✓ | | | ✓ | ✓ | ✓ |
| Trojan110 | ✓ | ✓ | ✓ | ✓ | | ✓ | | | ✓ | ✓ | ✓ |
| Trojan111 | | ✓ | ✓ | ✓ | | ✓ | | | ✓ | ✓ | ✓ |
| Trojan112 | | ✓ | ✓ | ✓ | | ✓ | | | ✓ | ✓ | ✓ |
| Trojan113 | | ✓ | ✓ | ✓ | | ✓ | | | ✓ | ✓ | ✓ |
| Trojan114 | | ✓ | ✓ | ✓ | | ✓ | | | ✓ | ✓ | ✓ |
| Trojan115 | | ✓ | ✓ | ✓ | | ✓ | | | ✓ | ✓ | ✓ |
| Trojan116 | | ✓ | ✓ | ✓ | | ✓ | | | ✓ | ✓ | ✓ |
| Trojan117 | ✓ | ✓ | ✓ | ✓ | | ✓ | | | ✓ | ✓ | ✓ |
| Trojan118 | | ✓ | ✓ | ✓ | | ✓ | | | ✓ | ✓ | ✓ |
| Trojan119 | | ✓ | ✓ | ✓ | | ✓ | | | ✓ | ✓ | ✓ |
| Trojan120 | | ✓ | ✓ | ✓ | | ✓ | | | ✓ | ✓ | ✓ |
| Trojan121 | | ✓ | ✓ | ✓ | | ✓ | | | ✓ | ✓ | ✓ |
| Trojan122 | | ✓ | ✓ | ✓ | | ✓ | | | ✓ | ✓ | ✓ |
| Trojan123 | | ✓ | ✓ | ✓ | | ✓ | | | ✓ | ✓ | ✓ |
| Trojan124 | | ✓ | ✓ | ✓ | | ✓ | | | ✓ | ✓ | ✓ |
| Trojan125 | | ✓ | ✓ | ✓ | | ✓ | | | ✓ | ✓ | ✓ |
| Trojan126 | | ✓ | ✓ | ✓ | | ✓ | | | ✓ | ✓ | ✓ |
| Trojan127 | | ✓ | ✓ | ✓ | | ✓ | | | ✓ | ✓ | ✓ |

**References**

[1] A. C. Kim, S. Kim, W. H. Park, and D. H. Lee, “Fraud and financial crime detection model using malware forensics,” *Multimedia Tools and Applications*, vol. 68, no. 2, pp. 479–496, Jan. 2014.

[2] M. Damshenas, A. Dehghantanha, and R. Mahmoud, “A Survey On Malware Propagation, Analysis, And Detection,” *International Journal of Cyber-Security and Digital Forensics (IJCSDF)*, vol. 2, no. 4, pp. 10–29, 2013.

[3] R. Pilling, “Global threats, cyber-security nightmares and how to protect against them,” *Computer Fraud and Security*, vol. 2013, no. 9, pp. 14–18.

[4] M. Riek, R. Bohme, and T. Moore, “Measuring the Influence of Perceived Cybercrime Risk on Online Service Avoidance,” *IEEE Transactions on Dependable and Secure Computing*, vol. 13, no. 2, pp. 261–273, Mar. 2016.

[5] H. Tiirmaa-Klaar, J. Gassen, E. Gerhards-Padilla, and P. Martini, “Botnets: How to Fight the Ever-Growing Threat on a Technical Level,” in *Botnets*, London: Springer London, 2013, pp. 41–97.

[6] M. Ficco, M. Choraś, and R. Kozik, “Simulation platform for cyber-security and vulnerability analysis of critical infrastructures,” *Journal of Computational Science*, Apr. 2017.

[7] E. I. Edem, C. Benzaid, A. Al-Nemrat, and P. Watters, “Analysis of Malware Behaviour: Using Data Mining Clustering Techniques to Support Forensics Investigation,” in *2014 Fifth Cybercrime and Trustworthy Computing Conference*, 2014, pp. 54–63.

[8] Y. Kim, I. Kim, and N. Park, “Analysis of Cyber Attacks and Security Intelligence,” in *Mobile, Ubiquitous, and Intelligent Computing*, vol. 274, Berlin, Heidelberg: Springer-Verlag Berlin Heidelberg, 2014, pp. 629–636.

[9] C. M. Colombini, A. Colella, M. Mattiucci, and A. Castiglione, “Cyber Threats Monitoring: Experimental Analysis of Malware Behavior in Cyberspace,” in *Lecture Notes in Computer Science (including subseries Lecture Notes in Artificial Intelligence and Lecture Notes in Bioinformatics)*, vol. 8128 LNCS, Berlin, Heidelberg: Springer Berlin Heidelberg, 2013, pp. 236–252.

[10] N. Lee, “Cyber Warfare: Weapon of Mass Disruption,” in *Counterterrorism and Cybersecurity*, Second, Ed. New York, NY: Springer New York, 2013, pp. 99–118.

[11] I. A. Saeed, J. B. Campus, M. A. Selamat, M. Ali, and M. A. Abuagoub, “A Survey on Malware and Malware Detection Systems,” *International Journal of Computer Applications*, vol. 67, no. 16, pp. 975–8887, 2013.

[12] Mohamad Fadli Zolkipli and A. Jantan, “An approach for malware behavior identification and classification,” in *Computer Research and Development (ICCRD), 2011 3rd International Conference*, 2011, pp. 191–194.

[13] M. F. Zolkipli and A. Jantan, “Malware Behavior Analysis: Learning and Understanding Current Malware Threats,” in *2010 Second International Conference on Network Applications, Protocols and Services*, 2010, pp. 218–221.

[14] N. Kiyavash, F. Koushanfar, T. P. Coleman, and M. Rodrigues, “A Timing Channel Spyware for the CSMA/CA Protocol,” *IEEE Transactions on Information Forensics and Security*, vol. 8, no. 3, pp. 477–487, Mar. 2013.

[15] A. Singh, B. Singh, and H. Joseph, "Malware Analysis," in *Vulnerability Analysis and Defense for the Internet*, Boston, MA: Springer US, 2008, pp. 169–211.

[16] D. Salomon, "Trojan Horses," in *Elements of Computer Security*, London: Springer-Verlag London Limited, 2010, pp. 123–135.

[17] V. S. Subrahmanian, M. Ovelgonne, T. Dumitras, and B. A. Prakash, *The Global Cyber-Vulnerability Report*, no. November 2013. Cham: Springer International Publishing, 2015.

[18] S. T. King and P. M. Chen, "SubVirt: implementing malware with virtual machines," in *2006 IEEE Symposium on Security and Privacy (S&P'06)*, 2006, p. 14 pp.-pp.327.

[19] L. Xianghe, Z. Liancheng, and L. Shuo, "Kernel rootkits implement and detection," *Wuhan University Journal of Natural Sciences*, vol. 11, no. 6, pp. 1473–1476, Nov. 2006.

[20] E. M. Hutchins, M. J. Clopp, and P. D. Rohan M. Amin, "Intelligence-driven computer network defense informed by analysis of adversary campaigns and intrusion kill chains," *Lockheed Martin Corporation*, no. July 2005, pp. 1–14, 2011.

[21] A. Buescher, F. Leder, and T. Siebert, "Banksafe Information Stealer Detection Inside the Web Browser," in *Recent Advances in Intrusion Detection*, vol. 6961 LNCS, Berlin, Heidelberg: Springer Berlin Heidelberg, 2011, pp. 262–280.

[22] C. Criscione, F. Bosatelli, S. Zanero, and F. Maggi, "ZARATHUSTRA: Extracting Webinject signatures from banking trojans," in *2014 Twelfth Annual International Conference on Privacy, Security and Trust*, 2014, pp. 139–148.

[23] K. E. Heckman, F. J. Stech, R. K. Thomas, B. Schmoker, and A. W. Tsow, *Cyber Denial, Deception and Counter Deception*. Cham: Springer International Publishing, 2015.

[24] M. Line, A. Zand, G. Stringhini, and R. Kemmerer, "Targeted Attacks against Industrial Control Systems: Is the Power Industry Prepared?," *Proceedings of the ACM Workshop on Smart Energy Grid Security*, pp. 13–22, 2014.

[25] M. S. Awan, P. Burnap, and O. F. Rana, "Estimating Risk Boundaries for Persistent and Stealthy Cyber-Attacks," *Proceedings of the 2015 Workshop on Automated Decision Making for Active Cyber Defense - SafeConfig '15*, pp. 15–20, 2015.

[26] J. Raphel and P. Vinod, "Heterogeneous Opcode Space for Metamorphic Malware Detection," *Arabian Journal for Science and Engineering*, vol. 42, no. 2, pp. 537–558, Feb. 2017.

[27] I. A. AL-Taharwa, H.-M. Lee, A. B. Jeng, K.-P. Wu, C.-H. Mao, T.-E. Wei, and S.-M. Chen, "RedJsod: A Readable JavaScript Obfuscation Detector Using Semantic-based Analysis," in *2012 IEEE 11th International Conference on Trust, Security and Privacy in Computing and Communications*, 2012, pp. 1370–1375.

[28] J. a P. Marpaung, M. Sain, and H.-J. Lee, "Survey on Malware Evasion Techniques: State of the Art and Challenges," in *14th International Conference on Advanced Communication Technology (ICACT)*, 2012, pp. 744–749.

[29] M. Alazab, S. Venkatraman, P. Watters, M. Alazab, and A. Alazab, "Cybercrime: The Case of Obfuscated Malware," in *Global Security, Safety and Sustainability & e-Democracy*, vol. 99 LNICST, Berlin, Heidelberg: Springer Berlin Heidelberg, 2012, pp. 204–211.

[30] E. W. Burger, M. D. Goodman, P. Kampanakis, and K. A. Zhu, "Taxonomy Model for Cyber Threat Intelligence Information Exchange Technologies," in *Proceedings of the 2014 ACM Workshop on Information Sharing & Collaborative Security - WISCS '14*, 2014, pp. 51–60.

[31] T. Yadav and A. M. Rao, "Technical Aspects of Cyber Kill Chain," in *Security in Computing and Communications*, vol. 377, Cham: Springer International Publishing Switzerland, 2015, pp. 438–452.

[32] S. Caltagirone, A. Pendergast, and C. Betz, "The Diamond Model of Intrusion Analysis," *Threat Connect*, vol. 298, no. 704, pp. 1–61, 2013.

[33] A. Al-Bataineh and G. White, "Analysis and detection of malicious data exfiltration in web traffic," in *2012 7th International Conference on Malicious and Unwanted Software*, 2012, pp. 26–31.

[34] L. Martin, "Cyber Kill Chain® · Lockheed Martin." [Online]. Available: http://www.lockheedmartin.com/us/what-we-do/aerospace-defense/cyber/cyber-kill-chain.html. [Accessed: 29-Mar-2017].

[35] R. Kozik and M. Choraś, "Current cyber security threats and challenges in critical infrastructures protection," *2013 2nd International Conference on Informatics and Applications, ICIA 2013*, pp. 93–97, 2013.

[36] S. Khattak, N. R. Ramay, K. R. Khan, A. A. Syed, and S. A. Khayam, "A Taxonomy of botnet behavior, detection, and defense," *IEEE Communications Surveys and Tutorials*, vol. 16, no. 2, pp. 898–924, 2014.

[37] N. Weaver, V. Paxson, S. Staniford, and R. Cunningham, "A taxonomy of computer worms," in *Proceedings of the 2003 ACM workshop on Rapid Malcode - WORM'03*, 2003, p. 11.

[38] G. Jacob, H. Debar, and E. Filiol, "Behavioral detection of malware: from a survey towards an established taxonomy," *Journal in Computer Virology*, vol. 4, no. 3, pp. 251–266, Aug. 2008.

[39] M. Lindorfer, C. Kolbitsch, and P. Milani Comparetti, "Detecting Environment-Sensitive Malware," in *Recent Advances in Intrusion Detection*, vol. 6961 LNCS, Berlin, Heidelberg: Springer Berlin Heidelberg, 2011, pp. 338–357.

[40] X. Ugarte-Pedrero, D. Balzarotti, I. Santos, and P. G. Bringas, "SoK: Deep Packer Inspection: A Longitudinal Study of the Complexity of Run-Time Packers," in *2015 IEEE Symposium on Security and Privacy*, 2015, vol. 2015–July, pp. 659–673.

[41] A. Karim, S. A. A. Shah, and R. Salleh, "Mobile Botnet Attacks: A Thematic Taxonomy," in *New Perspectives in Information Systems and Technologies*, vol. 2, Cham: Springer International Publishing, 2014, pp. 153–164.

[42] A. Amamra, C. Talhi, and J.-M. Robert, "Smartphone malware detection: From a survey towards taxonomy," in *2012 7th International Conference on Malicious and Unwanted Software*, 2012, pp. 79–86.

[43] L. Delosières and D. García, "Infrastructure for Detecting Android Malware," in *Information Sciences and Systems 2013*, vol. 264, Cham: Springer International Publishing Switzerland, 2013, pp. 389–398.

[44] D. Dagon, G. Gu, C. P. Lee, and W. Lee, "A Taxonomy of Botnet Structures," in *Twenty-Third Annual Computer Security Applications Conference (ACSAC 2007)*, 2007, pp. 325–339.

[45] B. B. Gupta, A. Tewari, A. K. Jain, and D. P. Agrawal, "Fighting against phishing attacks: state of the art and future challenges," *Neural Computing and Applications*, pp. 1–26, Mar. 2016.

[46] R. J. Rodríguez, "Evolution and characterization of point-of-sale RAM scraping malware," *Journal of Computer Virology and Hacking Techniques*, pp. 1–14, May 2016.

[47] O. Osanaiye, H. Cai, K.-K. R. Choo, A. Dehghantanha, Z. Xu, and M. Dlodlo, "Ensemble-based multi-

filter feature selection method for DDoS detection in cloud computing," *EURASIP Journal on Wireless Communications and Networking*, vol. 2016, no. 1, p. 130, Dec. 2016.

[48] F. Daryabar, A. Dehghantanha, and H. G. Broujerdi, "Investigation of malware defence and detection techniques," *International Journal of Digital Information and Wireless Communications (IJDIWC)*, vol. 1, no. 3, pp. 645–650, 2011.

[49] F. Daryabar, A. Dehghantanha, N. I. Udzir, and N. Fazlida, "Analysis of Known and Unknown Malware Bypassing Techniques," *International Journal of Information Processing and Management(IJIPM)*, vol. 4, no. 6, pp. 50–59, 2013.

[50] L. Rocha, "Malware Analysis – Dridex & Process Hollowing | Count Upon Security," 2015. [Online]. Available: https://countuponsecurity.com/2015/12/07/malware-analysis-dridex-process-hollowing/. [Accessed: 09-Mar-2017].

[51] T. K. Lengyel, S. Maresca, B. D. Payne, G. D. Webster, S. Vogl, and A. Kiayias, "Scalability, fidelity and stealth in the DRAKVUF dynamic malware analysis system," in *Proceedings of the 30th Annual Computer Security Applications Conference on - ACSAC '14*, 2014, pp. 386–395.

[52] T. K. Lengyel, "Malware Collection and Analysis via Hardware Virtualization," 2015.

[53] J. Leitch, "Process Hollowing," 2013.

[54] M. Z. Shafiq, S. A. Khayam, and M. Farooq, "Embedded Malware Detection Using Markov n-Grams," in *Detection of Intrusions and Malware, and Vulnerability Assessment*, Berlin, Heidelberg: Springer Berlin Heidelberg, 2008, pp. 88–107.

[55] F. Daryabar, A. Dehghantanha, and N. I. Udzir, "Investigation of bypassing malware defences and malware detections," in *2011 7th International Conference on Information Assurance and Security (IAS)*, 2011, pp. 173–178.

[56] D. Stevens, "Malicious PDF Documents Explained," *IEEE Security & Privacy Magazine*, vol. 9, no. 1, pp. 80–82, Jan. 2011.

[57] J. Song, J. Song, and J. Kim, "Detection of Heap-Spraying Attacks Using String Trace Graph," in *Information Security Applications*, vol. LNCS 8909, Cham: Springer International Publishing, 2015, pp. 17–26.

[58] N. Nissim, A. Cohen, and Y. Elovici, "Boosting the Detection of Malicious Documents Using Designated Active Learning Methods," in *2015 IEEE 14th International Conference on Machine Learning and Applications (ICMLA)*, 2015, pp. 760–765.

[59] C. Lin and C. Chen, "Efficient Spear-phishing Threat Detection Using Hypervisor Monitor," *The 49th Annual IEEE International Carnahan Conference on Security Technology*, pp. 299–303, 2011.

[60] A. Jayan and S. Dija, "Detection of spoofed mails," in *2015 IEEE International Conference on Computational Intelligence and Computing Research (ICCIC)*, 2015, pp. 1–4.

[61] J. Yu, E. Kim, H. Kim, and J. Huh, "A Framework for Detecting MAC and IP Spoofing Attacks with Network Characteristics," in *2016 International Conference on Software Security and Assurance (ICSSA)*, 2016, pp. 49–53.

[62] B. Liu, J. Bi, and A. V. Vasilakos, "Toward Incentivizing Anti-Spoofing Deployment," *IEEE Transactions on Information Forensics and Security*, vol. 9, no. 3, pp. 436–450, Mar. 2014.

[63] S. Gajek and A.-R. Sadeghi, "A Forensic Framework for Tracing Phishers," in *The Future of Identity in the Information Society*, Boston, MA: Springer US, 2008, pp. 23–35.

[64] I. R. A. Hamid and J. Abawajy, “Hybrid Feature Selection for Phishing Email Detection,” in *Algorithms and Architectures for Parallel Processing*, Springer, Berlin, Heidelberg, 2011, pp. 266–275.

[65] M.-E. Maurer and L. Höfer, “Sophisticated Phishers Make More Spelling Mistakes: Using URL Similarity against Phishing,” in *Cyberspace Safety and Security*, Springer, Berlin, Heidelberg, 2012, pp. 414–426.

[66] E. D. Frauenstein and R. von Solms, “An Enterprise Anti-phishing Framework,” in *Information Assurance and Security Education and Training*, Springer, Berlin, Heidelberg, 2013, pp. 196–203.

[67] J. Hajgude and L. Ragha, “‘Phish mail guard: Phishing mail detection technique by using textual and URL analysis,’” in *2012 World Congress on Information and Communication Technologies*, 2012, pp. 297–302.

[68] J. V. Chandra, N. Challa, and S. K. Pasupuleti, “A practical approach to E-mail spam filters to protect data from advanced persistent threat,” in *2016 International Conference on Circuit, Power and Computing Technologies (ICCPCT)*, 2016, pp. 1–5.

[69] K. Shaerpour, A. Dehghantanha, and R. Mahmod, “TRENDS IN ANDROID MALWARE DETECTION,” *Journal of Digital Forensics, Security and Law*, vol. 8, no. 3, pp. 21–40, 2013.

[70] U. P. Daniel and G. Epiphaniou, “Safeguarding Forensic Integrity of Virtual Environment Evidence,” *International Journal of Computer Applications*, vol. 82, pp. 975–8887, 2013.

[71] J. Walls and K. K. R. Choo, “A review of free cloud-based anti-malware apps for android,” in *Proceedings - 14th IEEE International Conference on Trust, Security and Privacy in Computing and Communications, TrustCom 2015*, 2015, vol. 1, pp. 1053–1058.

[72] S. Hosseini, M. A. Azgomi, and A. T. Rahmani, “Malware propagation modeling considering software diversity and immunization,” *Journal of Computational Science*, vol. 13, pp. 49–67, Mar. 2016.

[73] I. You and K. Yim, “Malware Obfuscation Techniques: A Brief Survey,” in *2010 International Conference on Broadband, Wireless Computing, Communication and Applications*, 2010, pp. 297–300.

[74] M. Musale, T. H. Austin, and M. Stamp, “Hunting for metamorphic JavaScript malware,” *Journal of Computer Virology and Hacking Techniques*, vol. 11, no. 2, pp. 89–102, May 2015.

[75] P. Vinod, V. Laxmi, M. S. Gaur, and G. Chauhan, “MOMENTUM: MetamOrphic malware exploration techniques using MSA signatures,” in *2012 International Conference on Innovations in Information Technology (IIT)*, 2012, pp. 232–237.

[76] G. Shanmugam, R. M. Low, and M. Stamp, “Simple substitution distance and metamorphic detection,” *Journal of Computer Virology and Hacking Techniques*, vol. 9, no. 3, pp. 159–170, Aug. 2013.

[77] S. Madenur Sridhara and M. Stamp, “Metamorphic worm that carries its own morphing engine,” *Journal of Computer Virology and Hacking Techniques*, vol. 9, no. 2, pp. 49–58, May 2013.

[78] S. Alam, I. Sogukpinar, I. Traore, and R. Nigel Horspool, “Sliding window and control flow weight for metamorphic malware detection,” *Journal of Computer Virology and Hacking Techniques*, vol. 11, no. 2, pp. 75–88, May 2015.

[79] M. Alazab, S. Venkataraman, and P. Watters, “Towards Understanding Malware Behaviour by the Extraction of API Calls,” in *2010 Second Cybercrime and Trustworthy Computing Workshop*, 2010, pp. 52–59.

[80] T. Barabosch, A. Dombeck, K. Yakdan, and E. Gerhards-Padilla, *Research in Attacks, Intrusions and*

*Defenses*, vol. 8688, no. 2012. Cham: Springer International Publishing, 2014.

[81] A. Karim, R. Bin Salleh, M. Shiraz, S. A. A. Shah, I. Awan, and N. B. Anuar, "Botnet detection techniques: review, future trends, and issues," *Journal of Zhejiang University SCIENCE C*, vol. 15, no. 11, pp. 943–983, Nov. 2014.

[82] A. K. Sood and S. Zeadally, "A Taxonomy of Domain-Generation Algorithms," *IEEE Security & Privacy*, vol. 14, no. 4, pp. 46–53, Jul. 2016.

[83] H.-K. Kang, J.-S. Kim, B.-I. Kim, and H.-C. Jeong, *IT Convergence and Security 2012*, vol. 215. Dordrecht: Springer Netherlands, 2013.

[84] D. Cosovan, R. Benchea, and D. Gavrilut, "A Practical Guide for Detecting the Java Script-Based Malware Using Hidden Markov Models and Linear Classifiers," in *2014 16th International Symposium on Symbolic and Numeric Algorithms for Scientific Computing*, 2014, pp. 236–243.

[85] D. O. Brien, "Dridex : Tidal waves of spam pushing dangerous financial Trojan," 2016. [Online]. Available: http://www.symantec.com/content/en/us/enterprise/media/security_response/whitepapers/dridex-financial-trojan.pdf. [Accessed: 11-Jul-2016].

[86] S. Mohtasebi and A. Dehghantanha, "A Mitigation Approach to the Privacy and Malware Threats of Social Network Services," in *Digital Information Processing and Communications*, Springer, Berlin, Heidelberg, 2011, pp. 448–459.

[87] C. K. Joe-Uzuegbu, U. C. Iwuchukwu, and L. C. Ezema, "Application virtualization techniques for malware forensics in social engineering," in *2015 International Conference on Cyberspace (CYBER-Abuja)*, 2015, pp. 45–56.

[88] P. Jyotiyana and S. Maheshwari, "A Literature Survey on Malware and Online Advertisement Hidden Hazards," in *Intelligent Systems Technologies and Applications 2016*, Springer, Cham, 2016, pp. 449–460.

[89] S. Gunalakshmii and P. Ezhumalai, "Mobile keylogger detection using machine learning technique," in *Proceedings of IEEE International Conference on Computer Communication and Systems ICCCS14*, 2014, pp. 051–056.

[90] S. Ortolani, C. Giuffrid, and B. Crispo, *Recent Advances in Intrusion Detection*, vol. 5758, no. 216917. Berlin, Heidelberg: Springer Berlin Heidelberg, 2009.

[91] P. Stewin, "Technical Background, Preliminaries and Assumptions," in *Detecting Peripheral-based Attacks on the Host Memory*, Cham: Springer International Publishing, 2015, pp. 9–19.

[92] P. Stewin and I. Bystrov, "Understanding DMA Malware," in *Detection of Intrusions and Malware, and Vulnerability Assessment*, vol. 7591 LNCS, Berlin, Heidelberg: Springer Berlin Heidelberg, 2013, pp. 21–41.

[93] F. L. Sang, E. Lacombe, V. Nicomette, and Y. Deswarte, "Exploiting an I/OMMU vulnerability," in *2010 5th International Conference on Malicious and Unwanted Software*, 2010, pp. 7–14.

[94] J. Choi, Y. Han, S. Cho, H. Yoo, J. Woo, M. Park, Y. Song, and L. Chung, "A Static Birthmark for MS Windows Applications Using Import Address Table," in *2013 Seventh International Conference on Innovative Mobile and Internet Services in Ubiquitous Computing*, 2013, pp. 129–134.

[95] A. Stewart, "DLL Side-Loading: A Thorn in the Side of the Anti-Virus Industry," *FireEye, Inc*, 2016. [Online]. Available: https://www.fireeye.com/content/dam/fireeye-www/global/en/current-threats/pdfs/rpt-dll-sideloading.pdf. [Accessed: 11-Jul-2016].

[96] B. Min and V. Varadharajan, “Secure Dynamic Software Loading and Execution Using Cross Component Verification,” in *2015 45th Annual IEEE/IFIP International Conference on Dependable Systems and Networks*, 2015, pp. 113–124.

[97] G. Yucheng, W. Peng, L. Juwei, and G. Qingping, “A Way to Detect Computer Trojan Based on DLL Preemptive Injection,” in *2011 10th International Symposium on Distributed Computing and Applications to Business, Engineering and Science*, 2011, pp. 255–258.

[98] D. D. Zovi, C. Eagle, I. Guilfanov, S. Porst, D. Quist, and P. Engbretson, *Practical Malware Analysis: The Hands-on Guide to Dissecting Malicious Software*. .

[99] G. Epiphaniou, T. French, H. Al-Khateeb, A. Dehghantanha, and H. Jahankhani, “A Novel Anonymity Quantification and Preservation Model for UnderNet Relay Networks,” Springer, Cham, 2016, pp. 371–384.

[100] K. Alieyan, A. ALmomani, A. Manasrah, and M. M. Kadhum, “A survey of botnet detection based on DNS,” *Neural Computing and Applications*, pp. 1–18, Dec. 2015.

[101] C. J. Dietrich, C. Rossow, F. C. Freiling, H. Bos, M. Van Steen, and N. Pohlmann, “On Botnets That Use DNS for Command and Control,” in *2011 Seventh European Conference on Computer Network Defense*, 2011, pp. 9–16.

[102] B. Donohue, “The Big Four Banking Trojans- Kaspersky Daily – Kaspersky Lab official blog,” 2013. [Online]. Available: https://www.kaspersky.com/blog/the-big-four-banking-trojans/2956/. [Accessed: 13-Aug-2017].

[103] R. Horne, “The cyber threat to banking Foreword,” *PWC*, 2014.

[104] A. Javed and M. Akhlaq, “Patterns in malware designed for data espionage and backdoor creation,” in *2015 12th International Bhurban Conference on Applied Sciences and Technology (IBCAST)*, 2015, pp. 338–342.

[105] P. Goyal, N. Bansal, and N. Gupta, “Averting man in the browser attack using user-specific personal images,” in *2013 3rd IEEE International Advance Computing Conference (IACC)*, 2013, pp. 1283–1286.

[106] F. Bin Mat Nor, K. Abd Jalil, and J. Ab Manan, “An enhanced remote authentication scheme to mitigate man-in-the-browser attacks,” in *Proceedings Title: 2012 International Conference on Cyber Security, Cyber Warfare and Digital Forensic (CyberSec)*, 2012, pp. 271–276.

[107] In-A Song and Young-Seok Lee, “Improvement of Key Exchange protocol to prevent Man-in-the-middle attack in the satellite environment,” in *2016 Eighth International Conference on Ubiquitous and Future Networks (ICUFN)*, 2016, pp. 408–413.

[108] M. Tiwari, T. Sharma, P. Sharma, S. Jindal, and Priyanshu, “Prevention of Man in the Middle Attack by Using Honeypot,” in *Proceedings of International Conference on Advances in Computing*, India: Springer India, 2013, pp. 593–600.

[109] G. Loukas, D. Gan, and Tuan Vuong, “A taxonomy of cyber attack and defence mechanisms for emergency management networks,” in *2013 IEEE International Conference on Pervasive Computing and Communications Workshops (PERCOM Workshops)*, 2013, no. March, pp. 534–539.

[110] A. Karim, R. Salleh, M. Shiraz, S. Shah, I. Awan, and N. Anuar, “Botnet detection techniques: review, future trends, and issues,” *Computers & Electronics*, vol. 15, no. 11, pp. 943–983, 2014.

[111] O. Fonseca, E. Fazzion, P. H. B Las-Casas, D. Guedes, W. Meira, C. Hoepers, K. Steding-Jessen, and

M. H. Chaves, “Neighborhoods and bands: an analysis of the origins of spam,” *Journal of Internet Services and Applications*, vol. 6, no. 1, p. 9, Dec. 2015.

[112] M. Takes, “Cascaded Simple Filters for Accurate and Lightweight Email-Spam Detection,” in *2010 Fourth International Conference on Emerging Security Information, Systems and Technologies*, 2010, pp. 160–165.

[113] M. Damshenas, A. Dehghantanha, K.-K. R. Choo, and R. Mahmud, “M0Droid: An Android Behavioral-Based Malware Detection Model,” *Journal of Information Privacy and Security*, vol. 11, no. 3, pp. 141–157, Jul. 2015.

[114] I. Firdausi, C. Lim, A. Erwin, and A. S. Nugroho, “Analysis of Machine learning Techniques Used in Behavior-Based Malware Detection,” in *2010 Second International Conference on Advances in Computing, Control, and Telecommunication Technologies*, 2010, pp. 201–203.

[115] M. Ramilli and M. Prandini, “Always the Same, Never the Same,” *IEEE Security & Privacy Magazine*, vol. 8, no. 2, pp. 73–75, Mar. 2010.

[116] D. B. Prelipcean, A. S. Popescu, and D. T. Gavrilut, “Improving Malware Detection Response Time with Behavior-Based Statistical Analysis Techniques,” in *2015 17th International Symposium on Symbolic and Numeric Algorithms for Scientific Computing (SYNASC)*, 2015, pp. 232–239.

[117] H. Haddad Pajouh, R. Javidan, R. Khayami, D. Ali, and K.-K. R. Choo, “A Two-layer Dimension Reduction and Two-tier Classification Model for Anomaly-Based Intrusion Detection in IoT Backbone Networks,” *IEEE Transactions on Emerging Topics in Computing*, pp. 1–1, 2016.

[118] H. H. Pajouh, G. Dastghaibyfard, and S. Hashemi, “Two-tier network anomaly detection model: a machine learning approach,” *Journal of Intelligent Information Systems*, vol. 48, no. 1, pp. 61–74, Feb. 2017.

[119] N. Milosevic, A. Dehghantanha, and K.-K. R. Choo, “Machine learning aided Android malware classification,” *International Journal of Computers & Electrical Engineering*, Feb. 2017.

[120] H. S. Galal, Y. B. Mahdy, and M. A. Atiea, “Behavior-based features model for malware detection,” *Journal of Computer Virology and Hacking Techniques*, vol. 12, no. 2, pp. 59–67, May 2016.

[121] Y. Kugisaki, Y. Kasahara, Y. Hori, and K. Sakurai, “Bot Detection Based on Traffic Analysis,” in *The 2007 International Conference on Intelligent Pervasive Computing (IPC 2007)*, 2007, pp. 303–306.

[122] A. Boukhtouta, S. A. Mokhov, N.-E. Lakhdari, M. Debbabi, and J. Paquet, “Network malware classification comparison using DPI and flow packet headers,” *Journal of Computer Virology and Hacking Techniques*, vol. 12, no. 2, pp. 69–100, May 2016.

[123] M. A. Ahmad, S. Woodhead, and D. Gan, “A countermeasure mechanism for fast scanning malware,” in *2016 International Conference On Cyber Security And Protection Of Digital Services (Cyber Security)*, 2016, pp. 1–8.

[124] T. Dargahi, A. Caponi, M. Ambrosin, G. Bianchi, and M. Conti, “A Survey on the Security of Stateful SDN Data Planes,” *IEEE Communications Surveys & Tutorials*, vol. PP, no. 99, pp. 1–1, 2017.

[125] D. Plohmann, E. Gerhards-Padilla, and F. Leder, “Botnets: Detection, Measurement, Disinfection & Defence,” *European Network and Information Security Agency (ENISA)*, p. 153, 2011.

[126] R. Zuech, T. M. Khoshgoftaar, and R. Wald, “Intrusion detection and Big Heterogeneous Data: a Survey,” *Journal of Big Data*, vol. 2, no. 1, p. 3, Dec. 2015.

[127] K. Thakur, S. Kopecky, M. Nuseir, M. L. Ali, and M. Qiu, “An Analysis of Information Security Event

Managers," in *2016 IEEE 3rd International Conference on Cyber Security and Cloud Computing (CSCloud)*, 2016, pp. 210–215.

[128] M. Ambrosin, M. Conti, and T. Dargahi, "On the Feasibility of Attribute-Based Encryption on Smartphone Devices," in *Proceedings of the 2015 Workshop on IoT challenges in Mobile and Industrial Systems - IoT-Sys '15*, 2015, pp. 49–54.

[129] P. Cascón, J. Ortega, Y. Luo, E. Murray, A. Díaz, and I. Rojas, "Improving IPS by network processors," *The Journal of Supercomputing*, vol. 57, no. 1, pp. 99–108, Jul. 2011.

[130] K.-H. Choi and D. Lee, "A study on strengthening security awareness programs based on an RFID access control system for inside information leakage prevention," *Multimedia Tools and Applications*, vol. 74, no. 20, pp. 8927–8937, Oct. 2015.

[131] M. Szczepanik and I. Jóźwiak, "Detecting New and Unknown Malwares Using Honeynet," in *Advances in Multimedia and Network Information System Technologies*, vol. 80, Berlin, Heidelberg: Springer Berlin Heidelberg, 2010, pp. 173–180.

[132] Lance Spitzner, "Honeypots: Are They Illegal? | Symantec Connect Community," 2003. [Online]. Available: https://www.symantec.com/connect/articles/honeypots-are-they-illegal. [Accessed: 02-Aug-2017].

[133] L. Christopher, K.-K. R. Choo, and A. Dehghantanha, "Honeypots for employee information security awareness and education training: A conceptual EASY training model," Jun. 2017.

[134] I. Technologies, "Honey pots, honey nets, and padded cell system." [Online]. Available: http://www.idc-online.com/technical_references/pdfs/data_communications/Honey_Pots_Honey_Nets_Padded_Cell_system.pdf. [Accessed: 09-Apr-2016].

[135] S. McCombie and J. Pieprzyk, "Winning the Phishing War: A Strategy for Australia," in *2010 Second Cybercrime and Trustworthy Computing Workshop*, 2010, pp. 79–86.

[136] J. White, J. S. Park, C. A. Kamhoua, and K. A. Kwiat, "Social network attack simulation with honeytokens," *Social Network Analysis and Mining*, vol. 4, no. 1, p. 221, Dec. 2014.

[137] S. Chakravarty, G. Portokalidis, M. Polychronakis, and A. D. Keromytis, "Detection and analysis of eavesdropping in anonymous communication networks," *International Journal of Information Security*, vol. 14, no. 3, pp. 205–220, Jun. 2015.